\RequirePackage{fix-cm}
\documentclass[twocolumn,epjc3]{svjour3}
\smartqed  
\RequirePackage{graphicx}
\usepackage{url}
\usepackage{subfigure}
\usepackage{graphicx,color,dcolumn}
\usepackage{url}
\usepackage{amsmath}
\RequirePackage{color}
\usepackage{color}

\newcommand{\cmd}{\color{blue}}
\newcommand {\brd}{\bf\cmd}

\begin{document}

\title{The Scattering of Dirac Spinors in Rotating Spheroids } \subtitle{}
\author{Zhi Fu, Gao\thanksref{addr1,addr2} Ci Xing, Chen$^{\dag}$\thanksref{e1, addr3}\, Na, Wang\thanksref{addr1} }

\thankstext{e1}{e-mail: daleccx@ustc.edu.cn}
\institute{Xinjiang Astronomical Observatory, CAS, 150, Science 1-Street, Urumqi, Xinjiang, 830011, China.
\label{addr1}
\and Key Laboratory of Radio Astronomy, CAS, Nanjing, Jiangshu, 210008, China
\label{addr2}
\and Department of Astronomy, University of Sciences and Technology of China, CAS, Hefei 230026, China.
\label{addr3}
}
\date{Received: date / Accepted: date}
\maketitle
\begin{abstract}
There are many stars that are rotating spheroids in the Universe, and studying them is of very important significance. Since the times of Newton, many astronomers and physicists have researched gravitational properties of stars by considering the moment equations derived from Eulerian hydrodynamic equations. In this paper we study the scattering of spinors of the Dirac equation, and in particular investigate the scattering issue in the limit case of rotating Maclaurin spheroids. Firstly we give the metric of a rotating ellipsoid star, then write the Dirac equation under this metric, and finally derive the scattering solution to the Dirac equation and establish a relation between differential scattering cross-section, $\sigma$, and stellar matter density, $\mu$. It is found that the sensitivity of $\sigma$ to the change in $\mu$ is proportional to the density $\mu$. Because of weak gravitational field and constant mass density, our results are reasonable. The results  can be applied to white dwarfs, main sequence stars, red giants, supergiant stars and so on, as long as their gravitational fields are so weak that they can be treated in the Newtonan approximations, and the fluid is assumed to be incompressible.
Notice that we take the star's matter density to be its average density and the star is not taken to be compact. Obviously our results cannot be used to study neutron stars and black holes.  In particular, our results are suitable for white dwarfs, which have average densities of about $10^{5}-10^{6}$\,g~cm$^{-3}$, corresponding to a range of mass of about $0.21-0.61 M_{\bigodot}$ and a range of radius of about $6000-10000$\,km.
\end{abstract}

\section{Introduction}
As we know, a physical model usually contains a set of most basic physical quantities, some of which can be used to describe the structure and evolution of matter, while others can be used to describe the motion state of an object. A mathematical model of the motion of matter is usually composed of fundamental equations, fundamental physical quantities and definite solution conditions such as initial conditions, boundary conditions and joint conditions. The research method of physics is to abstract the objective world into a physical model, and then convert it into a mathematical model. By solving the mathematical model, we can obtain the functional relationship between physical quantities and space-time coordinates, then shall examine the physical model by observation or experiment, and improve the physical model based on comparisons.

Assuming that three semi-axes of an elliposoid are $a, b$ and $c$, respectively, we define a spheroid as an elliposoid with $a=b>c$. As the most general physical model, a rotating  spheroid is composed of fluids (magnetohydrodynamic, turbulence and perfect fluid) with the pressure, density and temperature in a gravitational field, particle groups in the non-thermal equilibrium (distribution function), and boson and fermion fields. Especially for scalar fields, the fundamental equations are the Klein-Gordon equations in flat space-time and curved space-time, respectively\,\cite{b1},\cite{b2},\cite{b3},\cite{b4}. A theory of gravity will be tested by the gravitational observation effect in different celestial bodies. For the same theory of gravity, different coupling coefficients give different solutions to the basic equations. There may exist significant differences between different theories of gravity, e.g., the theory of gravity with torsion, superstring theory and supergravity\,\cite{b5}\,\cite{b6}\,\cite{b7},\cite{b8}.

In the nearly 80 years since  the Dirac spinors were defined, their roles in describing fermion fields have been well established and widely accepted. For convenience, let us review  the previous research work conducted by other authors related to spinors: Villalba and de Fisica ~\cite{b9} solved the massless Dirac equation in a non-stationary rotating causal Godel-type cosmological universe using the method of separation of variables; Dolan et al.~\cite{b10} studied the scattering of massive spin-half waves by a Schwarzschild black hole using analytical and numerical methods; Deleo \& Rotelli~\cite{b11} developed the potential scattering of a spinor within the context of perturbation field theory; In 2011, Stefano  et al.~\cite{b12} studied Dirac spinors in Bianchi type-I cosmological models within the framework of torsional $f(R)$-gravity,  while Bord\'{e} et al.~\cite{b13} presented a second-quantized field theory of massive particles with a spin of a half or antiparticle in the presence of a weak gravitational field treated as spin two external field in a flat Minkowski background. In 2012, Poplawski \cite{b14} made use of Einstein-Cartan-Sciama-Kibble\,(ECSK) theory of gravity to discuss non-singular, big-bounce cosmology from spinor-torsion coupling, and Daude and Kamran~\cite{b15}considered massive Dirac fields evolving in the exterior region of a 5-D Myers-Perry black hole and studied their propagation properties; In 2014, Brihaye et al.\cite{b16} studied Dirac equation for spherically symmetric space-time and application to a boson star in EGB gravity, and Ambrus \& Winstanley~\cite{b17} discussed Dirac fermions on an anti-desitter background; In 2015, Bini et al.~\cite{b18} discussed massless Dirac particles in the vacuum $C$-metric; In 2017, R\"{o}ken~\cite{b19} showed the separability of the massive Dirac equation in the non-extreme Kerr geometry in the horizon-penetrating advanced Eddington-Finkelstein coordinates; Dzhunushaliev \& Folomeev~\cite{b20} investigated Dirac star in the presence of Maxwell and Proca fields, and Oliveira~\cite{b21} investigated the influence of non-inertial and spin effects on the 2$-D$ Dirac oscillator interacting with a uniform magnetic field and with Aharonov-Bohm effects in the cosmic string space-time.

As we know, there are two kinds of methods to study Dirac equation describing fermions in the vicinity of different types black holes. One is the Newman-Penrose (NP) formalism, and another is direct methods. In 2016, Batic et al.~\cite{b22} first gave the Dirac equation in the Schwarzschild black hole metric to study the problem of embedded eigenvalues adopting the NP formalism s; In 2018, Kraniotis [3]~\cite{b23} mmade use of the same methods to derive the Dirac equation in the Kerr¨CNewman¨Cde Sitter (KNdS) black hole background using a generalized Kinnersley null tetrad, and the same year Blazquez-Salcedo and Knoll \& Knoll~\cite{b24} used the direct methods to study the massive Dirac equation in the near-horizon metric of the extremal five dimensional Myers-Perry black hole with equal angular momenta. Moreover, Dariescu et al.~\cite{b25} used the direct methods to obtain the Dirac equation in the background of the Garfinkle-Horowitz-Strominger black hole. In 2020, Ahmad et al.~\cite{b26} took advantage of the NP formalism to study the Dirac equation around a regular Bardeen black hole surrounded by quintessence. The above-mentioned articles are related to the Dirac equation in curved space-time.

The remainder of this paper is organized as follows. In Sec. 2, we present the calculation method. In Section 3, we define connections and give the derivations of related quantities. In Section 4, we derive the Dirac equation in the weak gravitational field approximation. In Sec. 5, we give the scattering solutions to the Dirac equation and discuss the scattering issue in a limit case of rotating Maclaurin spheroids, and summarize the whole paper and look forward to the future work in Sec. 6. In an appendix we give concrete expressions of the Dirac equation in the weak gravitational potential and write the matrix $M$ and the scattering solutions of the Dirac equation.

\section{Calculation method }
\label{sec:2}
Let us emphasize that there are two approaches to study gravitational properties:
(1) moment method, and (2)Dirac field scattering, particle geodesic motion and scalar field scattering.  We will use the second approach, because the advantage of this approach is that the gravitational properties are closely related to the observations, and a physical model will be constrained by observations.
In this paper, we will discuss the physical effect of a rotating spheroid: the scattering of Dirac spinors, which has never been done before. A specific method is first to give the metric of a rotating  spheroid, then write out the Dirac equation under the metric,and find the scattering solution, and finally give a relationship
between the scattering cross-section and the stellar density. According to the observed scattering amplitude, the density is expected to be determined to study the gravitational characteristics of spheroids, and the sensitivity of the scattering cross-section to the change of the density will be discussed.

In this paper, we will discuss the physical effect of a rotation spheroid: the scattering of Dirac spinors, which has never been done before. The specific method is to first give the metric of the spheroid, then write the Dirac equation under the metric, find the scattering solution, and finally give the relation between the scattering cross-section and the star density. According to the observed scattering amplitude, the gravitational properties of a rotating spheroid will be studied by determining the density, and the sensitivity of the scattering cross-section to the density change will be discussed.

Here, employing the quantum field theory of curved space-time, we will study the scattering of spinors of the Dirac equation by taking the following three steps: \\
(1)In the first step, we get the metric
\begin{equation}
\,\,\,\,\,\,\,\,\,\,g_{\mu\nu}, g^{\mu\nu}\Rightarrow e_{\mu}^{(a)}, e_{(b)}^{\nu},
\label{1}
\end{equation}
\begin{equation}
\,\,\,\,\,\,\,\,\,\Gamma_{\mu\nu}^{\lambda}\Rightarrow \omega_{\mu(b)}^{(a)}=-e_{(b)}^{\nu}(\partial_
{\mu}e_{\nu}^{(a)}-\Gamma_{\mu\nu}^{\lambda}e_{\lambda}^{(a)}).
\label{2}
\end{equation}
(2)In the second step, we will study the Dirac equation in a curved space-time (weak gravitational field)
\begin{equation}
\,\,\,\,\,\,\,\,\,\,i\gamma^{(c)} e_{(c)}^{\mu} D_{\mu} \Psi^{(a)}-m\Psi^{(a)}=0.
\label{3}
\end{equation}
where $a=0, 1, 2, 3$~\cite{b27}\cite{b28}\cite{b29}. The quantity $D_{\mu}\Psi^{(a)}$ is given as
\begin{equation}
\,\,\,\,\,\,\,\,\,D_{\mu}\Psi^{(a)}=\partial_{\mu}\Psi^{(a)}+\omega_{\mu(b)}^{(a)}
\Psi^{(b)}
\label{4}
\end{equation}
with\quad $\Psi^{(a)}=(\phi^{(a)}, \chi^{(a)})^{T}$.
For convenience, we replace $\gamma^{(c)}$ with $\gamma^{k(D)}$ , which is dependent on the Pauli matrix $\sigma_{k}$. Also we get the second term on the right side of  Eq.\,(4), namely
\begin{equation}
\,\,\,\,\,\,\,\,\,\,\,\,\,\,\,\,\,\,\,\Pi_{\mu}^{(a)}\equiv\omega_{\mu(b)}^{(a)}\Psi^{(b)},
\label{5}
\end{equation}
where the summation convention is used. At the same time, the term of $\gamma^{(c)}e_{(c)}^{\mu}$ will be calculated, then we obtain the term of $\gamma^{(c)}e_{(c)}^{\mu}D_{\mu}\Psi^{(b)}$. Therefore, we can investigate the Dirac equation in curved space-time.

Rearranging the above Dirac equation leads to
\begin{equation}
\,\,\,\,\,\,\,\,\,\,\,\,\,\partial_{0}\Pi-\vec{\sigma_{k}}\cdot \bigtriangledown\Pi+P\Pi=0,
\label{6}
\end{equation}
For the physical meanings of quantities in Eq\,(6), to see the following sections.
(3)In the last step, we will obtain the scattering solution of the Dirac equation in a curved space-time i.e.,
\begin{small}
\begin{eqnarray}
\Psi(\vec{x})=\varphi(\vec{x})-\lim\limits _{\varepsilon\rightarrow 0}\int\limits _{-\infty} ^{\infty}\int\limits _{-\infty} ^{\infty}\int\limits _{-\infty} ^{\infty}d^{3}\acute{x}G_{\varepsilon}^
{\pm}(\vec{x},\acute{\vec{x}})V(\acute{\vec{x}})\Psi(\acute{\vec{x}}),
\label{7}
\end{eqnarray}
\end{small}
where the Green function $G_{\varepsilon}^{\pm}(\vec{x},\acute{\vec{x}})
=\frac{1}{H_{0}-E\pm i\varepsilon}\delta(\vec{x}-\acute{\vec{x}})$, with the first-order Hamiltonian
\begin{equation}
 \,\,\,\,\,\,\,\,\,\,\,\,\,\,\,H_{0}=-\vec{\sigma_{k}}\cdot\bigtriangledown+P_{0}\pm i\varepsilon.
 \label{8}
\end{equation}
Using the first-order weak gravitational field approximation, the $\Psi(\acute{\vec{x}})$ on the right side of Eq.\,(4) is replaced by the solution of free particle equation, and the scattering solution is obtained.
\section{Connections and related quantities}
\label{sec:3}
The Newtonian limit corresponds to the following line element~\cite{b30}
\begin{small}
\begin{eqnarray}
ds^{2}&&=(1+2\Phi)(dx^{0})^{2}-(1-2\Phi)[(dx^{1})^{2}+(dx^{2})^{2}  \nonumber\\
&& +(dx^{3})^{2}]\equiv\eta_{(a)(b)}\omega^{(a)}\omega^{(b)},~\omega^{(a)}=e_{\mu}^{(a)}dx^{\mu}.
\label{9}
\end{eqnarray}
\end{small}
From Eq.\,(9), it is obtained that
\begin{eqnarray}
&&~~~e_{\mu}^{(a)}=diag (1+\Phi,1-\Phi,1-\Phi,1-\Phi),\nonumber\\
&&~~~e_{(b)}^{\nu}= diag (1-\Phi,1+\Phi,1+\Phi,1+\Phi).
\label{10}
\end{eqnarray}
 According to the metric above, we get nonvanishing connections of $\Gamma_{\mu\nu}^{\lambda}$ \,($\mu, \nu, \lambda=0, 1, 2, 3$, and there is no summation over repeated indices)\\
(1)Case of $\Gamma_{\mu\nu}^{0}$
\begin{small}
\begin{eqnarray}
&& ~~\Gamma_{0i}^{0}=\Gamma_{i0}^{0}=(1-2\Phi)\frac{\partial \Phi}{\partial x^{i}},~~(i=1,2,3);
\label{11}
\end{eqnarray}
\end{small}
(2)Case of $\Gamma_{\mu\nu}^{j} (j=1,2,3)$
\begin{small}
\begin{eqnarray}
&& ~~\Gamma_{ij}^{j}=\Gamma_{ji}^{j}=-(1+2\Phi)\frac{\partial\Phi}{\partial x^{i}}~~~~~(i,j=1,2,3),\nonumber\\
&& ~~ \Gamma_{ii}^{j}=(1+2\Phi)\frac{\partial\Phi}{\partial x^{j}}~~~~~(i,j=1,2,3, i\neq j);
\label{12}
\end{eqnarray}
\end{small}
 Based on Eq.\,(2), the quantities\,$\omega_{\mu(b)}^{(a)}$ can be calculated as follows. In the following cases of (1-4), we set $a, b, c=1, 2, 3$, and there is no summation over the indices\\
(1)Case of $\omega_{0(0)}^{(0)}, \omega_{0(b)}^{(0)}, \omega_{0(0)}^{(a)}$ and $\omega_{0(b)}^{(a)}$ :
\begin{small}
\begin{eqnarray}
&&~~~\omega_{0(b)}^{(0)}=(1+\Phi)^{2}(1-2\Phi)\frac{\partial\Phi}{\partial x^{b}},\nonumber\\
&& ~~~~\omega_{0(0)}^{(a)}=(1-\Phi)^{2}(1+2\Phi)\frac{\partial\Phi}{\partial x^{a}},
\omega_{0(0)}^{(0)}=\omega_{0(b)}^{(a)}=0;
\label{13}
\end{eqnarray}
\end{small}
(2)Case of $\omega_{c(0)}^{(0)}$;
\begin{small}
\begin{eqnarray}
 ~~~~\omega_{c(0)}^{(0)}=-(1-\Phi)\frac{\partial\Phi}{\partial x^{c}}+(1-\Phi^{2})(1-2\Phi)\frac{\partial\Phi}{\partial x^{c}};
\label{14}
\end{eqnarray}
\end{small}
(3)Case of $\omega_{c(b)}^{(0)}$ and $\omega_{c(0)}^{(a)}$:
\begin{small}
\begin{eqnarray}
&&~~~\omega_{c(b)}^{(0)}=-(1+\Phi)\frac{\partial\Phi}{\partial x^{c}}, \omega_{c(0)}^{(a)}=(1-\Phi)\frac{\partial\Phi}{\partial x^{c}};
\label{15}
\end{eqnarray}
\end{small}
(4)Case of $\omega_{c(b)}^{(a)}$:
\begin{small}
\begin{eqnarray}
&&~~\omega_{c(a)}^{(a)}=(1+\Phi)\frac{\partial\Phi}{\partial x^{c}}-(1-\Phi^{2})(1+2\Phi)\frac{\partial\Phi}{\partial x^{c}}, \nonumber\\
&&~~~ \omega_{c(c)}^{(a)}=(1+\Phi)\frac{\partial\Phi}{\partial x^{c}}+ (1-\Phi^{2})(1+2\Phi)
\frac{\partial\Phi}{\partial x^{a}},\,(a\neq c), \nonumber\\
&&~~ \omega_{a(b)}^{(a)}=(1+\Phi)\frac{\partial\Phi}{\partial x^{a}}-(1-\Phi^{2})(1+2\Phi)\frac{\partial\Phi}{\partial x^{b}},\,(a\neq b) \nonumber\\
&&~~~\omega_{c(b)}^{(a)}=\omega_{c(a)}^{(b)}
=(1+\Phi)\frac{\partial\Phi}{\partial x^{c}},\,(a \neq b \neq c).
\label{16}
\end{eqnarray}
\end{small}
\section{The Dirac equation in a weak gravitational field approximation}
\label{sec:4}
As shown in Section 2, the Dirac equation in curved space-time in a weak gravitational field
approximation can be described by Eq.\,(3) and Eq.\,(4), with the wave function $\Psi^{(a)}=(\phi^{(a)},\chi^{(a)})^{T}$, $\phi^{(a)}=(\phi_{1}^{(a)},\phi_{2}^{(a)})^{T}$ and $\chi^{(a)}=(\chi_{1}^{(a)},\chi_{2}^{(a)})^{T}$, where $a=0, 1, 2, 3$. There are also mounting concerns that $m\Psi^{(a)}=m
\begin{pmatrix}
\phi^{(a)}\\
\chi^{(a)}
\end{pmatrix}$ and
\begin{small}
\begin{eqnarray}
\quad
D_{\mu}\Psi^{(a)}&&=D_{\mu}
\begin{pmatrix}
\varphi^{(a)}\\
\chi^{(a)}
\end{pmatrix}
=\begin{pmatrix}
\partial_{\mu}\phi^{(a)}\\
 \partial_{\mu}\chi^{(a)}
\end{pmatrix}
+\omega_{\mu(b)}^{(a)}
\begin{pmatrix}
\phi^{(b)}\\
\chi^{(b)}
\end{pmatrix}.
\label{17}
\end{eqnarray}
\end{small}
For convenience, by making the following substitution
\begin{small}
\begin{eqnarray}
&&\gamma^{(c)}\rightarrow\gamma^{k(D)},\gamma^{0(D)}=
\begin{pmatrix}
 I&0&\\
 0&-I&
\end{pmatrix},
  \gamma^{k(D)}=
\begin{pmatrix}
 0&\sigma_{k}&\\
-\sigma_{k}&0&
\end{pmatrix},
\label{18}
\end{eqnarray}
\end{small}
it is easy to obtain $\gamma^{(c)}e_{(c)}^{0}=
\begin{pmatrix}
I&0\\
0&-I
\end{pmatrix}
$ and $\gamma^{(c)}e_{(c)}^{k}=
\begin{pmatrix}
0&\sigma_{k}\\
-\sigma_{k}&0
\end{pmatrix}$, here $k=1, 2, 3$, and $\sigma_{k}$ is the Pauli matrix \footnote{In quantum mechanism the three components of the Pauli matrix under the $2\times2$ spin representation are
$ \sigma_{1}=
\begin{pmatrix}
0&1\\
1&0
\end{pmatrix},\,
\sigma_{2}=
\begin{pmatrix}
0&-i\\
i&0
\end{pmatrix},\,
{\rm and}
\sigma_{3}=
\begin{pmatrix}
1&0\\
0&-1
\end{pmatrix}.$}
Notice that we reconsider Eq.\,(17), and the second term on its right side is denoted as
$ \Pi_{\mu}^{(a)}=\omega_{\mu(b)}^{(a)}
\begin{pmatrix}
\varphi^{(b)}\\
\chi^{(b)}
\end{pmatrix},~\mu=0, 1, 2, 3.
$
The concrete values of $\Pi_{\mu}^{(a)}$ are written in a more compact way as follows \\
\begin{small}
\begin{eqnarray}
&&~\Pi_{0}^{(0)}=\sum_{b=1}^{3}\frac{\partial\Phi}{\partial x^{b}}\Psi^{(b)},~~ \Pi_{0}^{(a)}=\frac{\partial\Phi}{\partial x^{a}}\Psi^{(0)},\nonumber\\
&&~\Pi_{c}^{(0)}=-\Phi\frac{\partial\Phi}{\partial x^{c}}\Psi^{(0)}-\sum_{b=1}^{3}(1+\Phi)\frac{\partial\Phi}{\partial x^{c}}\Psi^{(b)},\nonumber\\
&&~\Pi_{a}^{(a)}=(1-\Phi)\frac{\partial\Phi}{\partial x^{a}}\Psi^{(0)}+
\sum_{b=1}^{3}[(1+\Phi)\frac{\partial\Phi}{\partial x^{a}}\nonumber\\
&&~~~~-(1+2\Phi)\frac{\partial\Phi}{\partial x^{b}}]\Psi^{(b)}(b\neq a)-\Phi\frac{\partial\Phi}{\partial x^{a}} \Psi^{(a)},\nonumber\\
&&~~\Pi_{c}^{(a)}(c\neq a)=(1-\Phi)\frac{\partial\Phi}{\partial x^{c}}\Psi^{(0)}\nonumber\\
&&~~+[(1+\Phi)\frac{\partial\Phi}{\partial x^{c}}+(1+2\Phi)\frac{\partial\Phi}{\partial x^{a}}]\Psi^{(c)}
 \nonumber\\
&&~~~+(1+\Phi)\frac{\partial\Phi}{\partial x^{c}}\Psi^{(b)}(b\neq a\neq c)
-\Phi\frac{\partial\Phi}{\partial x^{c}}\Psi^{(a)}.
\label{19}
\end{eqnarray}
\end{small}
(Here, we don't adopt Einstein convention in order to avoid confusion.)\\
In order to study the Dirac equation in curved space-time, we also calculate the quantity of $i\gamma^{(c)}e_{(c)}^{\mu}D_{\mu}\Psi^{(a)}$
\begin{small}
\begin{eqnarray}
&&  i\gamma^{(c)}e_{(c)}^{\mu}D_{\mu}\Psi^{(b)}\nonumber\\
&& ~~~=i
\begin{pmatrix}
I&0\\
0&-I
\end{pmatrix}
(1-\Phi)[\partial_{0}
\begin{pmatrix}
\phi^{(b)}\\
\chi^{(b)}
\end{pmatrix}
+\Pi_{0}^{(b)}]\nonumber\\
&& ~~~+i\sigma_{i}
\begin{pmatrix}
0&I\\
-I&0
\end{pmatrix}
(1+\Phi)[\partial_{i}
\begin{pmatrix}
\varphi^{(b)}\\
\chi^{(b)}
\end{pmatrix}
+\Pi_{i}^{(b)}],
\label{20}
\end{eqnarray}
\end{small}
where $b=0,1,2,3$ and the Einstein convention is used.

From the analysis above, the concrete expressions of the Dirac equation in a weak gravitational potential are derived, as shown in Appendix A. For convenience, Eq.(A1)-(A5) are expressed in a uniform and compact form:
\begin{equation}
\quad  \quad \quad \partial_{0}\Pi-\vec{\sigma}_{k}\cdot\bigtriangledown\Pi+P\Pi=0,
\label{21}
\end{equation}
where $\Pi=(\phi^{(0)},\,\chi^{(0)},\,\phi^{(1)},\,\chi^{(1)},\, \phi^{(2)},\,\chi^{(2)},\,\phi^{(3)},\, \chi^{(3)})^{T}$, $\sigma_{k}$\, is the Pauli matrix and $P$ is a $[8\times8]$ matrix, which is related to gravitational potential, the matrix elements are given as following:

1. For the first row,
\begin{small}
\begin{eqnarray}
&&~~~~ P_{11}=im,~P_{12}=0,~P_{13}=\frac{\partial\Phi}{\partial x},\nonumber\\
&& ~~~~P_{14}=P_{16}=P_{18}=-\frac{\partial\Phi}{\partial x^{i}}\sigma_{i}\,(i=1,2,3), \nonumber\\
&& ~~~~P_{15}=\frac{\partial\Phi}{\partial y},~~P_{17}=\frac{\partial\Phi}{\partial z};
\label{22}
\end{eqnarray}
\end{small}
2. For the second row,
\begin{small}
\begin{eqnarray}
&& ~~~~P_{21}=0,~~P_{22}=-im,\nonumber\\
&&~~~~P_{23}=P_{25}=P_{27}=-\frac{\partial\Phi}{\partial x^{i}}\sigma_{i}\,(i=1,2,3),\nonumber\\
&&~~~~P_{24}=-\frac{\partial\Phi}{\partial x},P_{26}=\frac{\partial\Phi}{\partial y},P_{24}=\frac{\partial\Phi}{\partial z};
\label{23}
\end{eqnarray}
\end{small}
3. For the third row,
\begin{small}
\begin{eqnarray}
&&~~~~P_{31}=\frac{\partial\Phi}{\partial x}, P_{32}=\frac{\partial\Phi}{\partial x^{i}}\sigma_{i} (i=1,2,3), \nonumber\\
&&~~~~P_{36}=\frac{\partial\Phi}{\partial x^{i}}\sigma_{i}+\varepsilon_{3ij}
(\frac{\partial\Phi}{\partial x^{i}}\sigma_{j}),\nonumber \\
&& ~~~~P_{38}=\frac{\partial\Phi}{\partial x^{i}}\sigma_{i}-\varepsilon_{2ij}
(\frac{\partial\Phi}{\partial x^{i}}\sigma_{j}),\nonumber\\
&&~~~~P_{33}=im,~P_{34}= P_{35}=P_{37}=0;
\label{24}
\end{eqnarray}
\end{small}
4. For the fourth row,
\begin{small}
\begin{eqnarray}
&& ~~~P_{42}=\frac{\partial\Phi}{\partial x},~P_{41}=\frac{\partial\Phi}{\partial x^{i}}\sigma_{i}~(i=1,2,3),~\nonumber\\
&&~~~~~P_{45}=P_{36}=\frac{\partial\Phi}{\partial x^{i}}\sigma_{i}+\varepsilon_{3ij}
(\frac{\partial\Phi}{\partial x^{i}}\sigma_{j}),\nonumber\\
&& ~~~~~P_{47}=P_{38}=\frac{\partial\Phi}{\partial x^{i}}\sigma_{i}-\varepsilon_{2ij}
(\frac{\partial\Phi}{\partial x^{i}}\sigma_{j}), \nonumber\\
&&~~~~~P_{43}=P_{46}=P_{48}=0,~ P_{44}=-im;
\label{25}
\end{eqnarray}
\end{small}
5. For the fifth row,
\begin{small}
\begin{eqnarray}
&& ~~~~P_{51}=\frac{\partial\Phi}{\partial y}, ~P_{52}=\frac{\partial\Phi}{\partial x^{i}}\sigma_{i}~(i=1,2,3),\nonumber\\
&&~~~~P_{54}=\frac{\partial\Phi}{\partial x^{i}}\sigma_{i}-\varepsilon_{3ij}
(\frac{\partial\Phi}{\partial x^{i}}\sigma_{j}),
\nonumber\\
&&~~~~P_{58}=\frac{\partial\Phi}{\partial x^{i}}\sigma_{i}+\varepsilon_{1ij}
(\frac{\partial\Phi}{\partial x^{i}}\sigma_{j}),\nonumber\\
&&~~~~P_{55}=im,~~P_{56}=P_{57}=P_{53}=0;
\label{26}
\end{eqnarray}
\end{small}
6. For the sixth row,
\begin{small}
\begin{eqnarray}
&& ~~~~~~P_{62}=\frac{\partial\Phi}{\partial y},~P_{61}=\frac{\partial\Phi}{\partial x^{i}}\sigma_{i}~(i=1,2,3), \nonumber \\
&&~~~~~P_{63}=P_{54}=\frac{\partial\Phi}{\partial x^{i}}\sigma_{i}-\varepsilon_{3ij}
(\frac{\partial\Phi}{\partial x^{i}}\sigma_{j}),\nonumber\\
&& ~~~~~P_{67}=P_{58}=\frac{\partial\Phi}{\partial x^{i}}\sigma_{i}+\varepsilon_{1ij}
(\frac{\partial\Phi}{\partial x^{i}}\sigma_{j}),\nonumber\\
&& ~~~~~P_{64}=P_{65}=P_{68}=0,~ P_{66}=-im;
\label{27}
\end{eqnarray}
\end{small}
7. For the seventh row,
\begin{small}
\begin{eqnarray}
&&~~~~~P_{71}=\frac{\partial\Phi}{\partial z},~ P_{72}=\frac{\partial\Phi}{\partial x^{i}}\sigma_{i}~(i=1,2,3),\nonumber\\
&& ~~~~~P_{74}=\frac{\partial\Phi}{\partial x^{i}}\sigma_{i}+\varepsilon_{2ij}
(\frac{\partial\Phi}{\partial x^{i}}\sigma_{j}),\nonumber \\
&& ~~~~~P_{76}=\frac{\partial\Phi}{\partial x^{i}}\sigma_{i}-\varepsilon_{1ij}
(\frac{\partial\Phi}{\partial x^{i}}\sigma_{j}),\nonumber\\
&&~~~~~P_{73}=P_{75}=P_{78}=0,~P_{76}=im;
\label{28}
\end{eqnarray}
\end{small}
8. For the eighth row,
\begin{small}
\begin{eqnarray}
&&~~~~~ P_{82}=\frac{\partial \Phi}{\partial z},~P_{81}=\frac{\partial\Phi}{\partial x^{i}}\sigma_{i}~(i=1,2,3),\nonumber\\
&&~~~~~P_{83}=P_{74}=\frac{\partial\Phi}{\partial x^{i}}\sigma_{i}+\varepsilon_{2ij}
(\frac{\partial\Phi}{\partial x^{i}}\sigma_{j}),\nonumber\\
&&~~~~~P_{85}=P_{76}=\frac{\partial\Phi}{\partial x^{i}}\sigma_{i}-\varepsilon_{1ij}
(\frac{\partial\Phi}{\partial x^{i}}\sigma_{j}),\nonumber\\
&&~~~~~P_{84}=P_{86}=P_{87}=0,~~P_{88}=-im,
\label{29}
\end{eqnarray}
\end{small}
 where $\varepsilon_{kij}$ is the Levi-Civita symbol.

By defining the matrix of $P_{*}\equiv P-P_{0}$ that is a small quantity, we get $\partial_{0}\Pi+H\Pi=0,$ and
\begin{eqnarray}
~~~~~~H=H_{0}+P_{*}=(-\vec{\sigma}_{k}\cdot\nabla+P_{0})+P_{*},
\label{30}
\end{eqnarray}
where $P_{0}$=diag$(im, -im, im, -im, im, -im, im, -im)$.

In order to get the scattering solution to the above Dirac equation using a perturbation theory, for convenience, we first set $\Phi=\bigwedge e^{iwt},\,(H_{0}-E)\bigwedge=-P_{*}\bigwedge$ and make the substitutions of $\bigwedge\rightarrow|\Psi>$, $P_{*}\rightarrow V$ and $iw\rightarrow E$, then we obtain $(H_{0}-E)|\Psi> =-V|\Psi>$ and $H_{0}|\varphi>=E^{(0)}|\varphi>$.
\section{The scattering solution to the Dirac equation and rotating Maclaurin spheroids}
\label{sec:5}
The main research objects of quantum mechanics are divided into two types: bound states and scattering states. Theoretically, the scattering state is a non-bound state, which involves the continuous region of the energy spectrum of a system.
One can freely control the energy of incident particles, which is different from dealing with particles in the bound state. The bound state theory mainly involves the eigenvalues and eigenstates of discrete, quantized energies of the system\,\cite{b31},\,\cite{b32},\,\cite{b33}.. The scattering theory mainly deals with the redistribution of scattering particles and their properties (such as polarization, correlation, etc.) in the scattering process. By analyzing the scattering results, one can find the structures inside particles, which promotes the development of basic theories. From trapped atoms to liberated quarks, a better understanding of the structure of matter depends largely on the study of scattering\,\cite{b34}.
\subsection{The scattering solution to the Dirac equation in a weak gravitational field of a rotating spheroid}
In this subsection, at first, we focus on a solution to the equation of Dirac spinors of free particles,
 $|\varphi>=|\varphi_{0}>e^{-i\vec{k}\cdot\vec{r}}$, which satisfies the following equation
  \begin{equation}
  ~~~~~~~~~(\vec{\sigma}_{k}\cdot i\vec{k}+P_{0})|\varphi_{0}>=E^{(0)}|\varphi_{0}>.
  \label{31}
  \end{equation}
From the above expression, it is obvious that the secular equation is written as
\begin{small}
\begin{eqnarray}
[{\rm det}(\vec{\sigma}_{k}\cdot i\vec{k}+im-E^{(0)}){\rm det}(\vec{\sigma}_{k}\cdot i\vec{k}-im-E^{(0)})]^{4}=0,
\label{32}
\end{eqnarray}
\end{small}
where
\begin{small}
\begin{eqnarray}
 ~~~~~~~~\vec{\sigma}_{k}\cdot i\vec{k}
&& =\sigma_{1}ik_{x}+\sigma_{2}ik_{y}+\sigma_{3}ik_{z}\nonumber\\
&&~~=
\small\begin{pmatrix}
  ik_{z}&ik_{x}+k_{y} \\
  ik_{x}-k_{y}&-ik_{z}\\
\end{pmatrix}.
\label{33}
\end{eqnarray}
\end{small}\\
There are four different solutions to Eq.\,(32): $E^{(0)}=im\pm ik$ and $E^{(0)}=-im\pm ik$. As the space is limited, we only choose the solution of $E^{(0)}=im+ik$. It should be noted that the choice of
 $E^{0}$ depends on the observation of the scattering cross section\,(see the end of this section). Supposing $k_x=k_y=0$ and $k=k_z$ and solving the above eigenvalues, we obtain
 \begin{small}
\begin{eqnarray}
\quad \quad &&\phi^{(0)}=\phi^{(1)}=\phi^{(2)}=\phi^{(3)}= e^{ik_{z}z}
\begin{pmatrix}
1\\
0\\
\end{pmatrix},\nonumber\\
&&~~~~\chi^{(0)}=\chi^{(1)}=\chi^{(2)}=\chi^{(3)}=0.
\label{34}
\end{eqnarray}
\end{small}
According to the scattering formula in quantum mechanics, we have
\begin{small}
\begin{eqnarray}
~~~~\Psi(\vec{x})&&\approx\varphi(\vec{x})-\lim\limits_{\varepsilon\rightarrow 0}\int d^{3} x'G^{\pm}_{\varepsilon}(\vec{x},\vec{x}')V(\vec{x}')\varphi(\vec{x}'),\nonumber\\
&&~~~~G^{\pm}_{\varepsilon}(\vec{x},\vec{x}')=\frac{1}{H_{0}-E\pm i\varepsilon}\delta(\vec{x},\vec{x}')\nonumber\\
&&~~~~~=\frac{1}{-\vec{\sigma}_{k}\cdot\nabla+P_{0}-E\pm i\varepsilon}\delta(\vec{x},\vec{x}'),
\label{35}
\end{eqnarray}
\end{small}
Based on the expressions of $\sigma_{\vec{k}}\cdot\nabla=
\begin{pmatrix}
\partial_{z}&\partial_{x}-i\partial_{y}\\
\partial_{x}+i\partial_{y}&-\partial_{z}\\
\end{pmatrix}$ and $\delta(\vec{x}-\vec{x}')=\frac{1}{(2\pi)^{3}}\int d^{3} pe^{-i\vec{p}\cdot(\vec{x}-\vec{x}')}$,
we first set $\varepsilon\rightarrow 0$, and define the matrices of $H$, $H_{01}$, and $H_{02}$ as follows:
\begin{eqnarray}
&&~~~~~~~~~H=-\vec{\sigma}_{k}\cdot\nabla+P_{0}-E\pm i\varepsilon,\nonumber\\
&&~~~~~~~~~~~H_{01}=-\vec{\sigma}_{k}\cdot\nabla+im-E\pm i\varepsilon, \nonumber\\
&&~~~~~~~~~~~H_{02}=-\vec{\sigma}_{k}\cdot\nabla-im-E\pm i\varepsilon.
\label{36}
\end{eqnarray}
Then we have
\begin{small}
\begin{eqnarray}
&&~~~~H_{01}=
\begin{pmatrix}
ip_{z}+im-E\pm i\varepsilon & ip_{x}+p_{y}\\
ip_{x}-p_{y} & -ip_{z}+im-E\pm i\varepsilon\\
\end{pmatrix}\nonumber\\
&&~~~~H_{02}=
\begin{pmatrix}
ip_{z}-im-E\pm i\varepsilon & ip_{x}+p_{y}\\
ip_{x}-p_{y} & -ip_{z}-im-E\pm i\varepsilon\\
\label{37}
\end{pmatrix}.
\end{eqnarray}
\end{small}
 Accordingly, their inverse matrices are written as
\begin{small}
\begin{eqnarray}
~~&&H_{01}^{-1}=\frac{1}{(im-E\pm i\varepsilon)^{2}+p^{2}}\times\nonumber\\
&&~~~~\begin{pmatrix}
-ip_{z}+im-E\pm i\varepsilon & -ip_{x}+p_{y}\\
-ip_{x}-p_{y} & ip_{z}+im-E\pm i\varepsilon\\
\end{pmatrix} \nonumber\\
&&~~H_{02}^{-1}=\frac{1}{(im-E\pm i\varepsilon)^{2}+p^{2}}\times\nonumber\\
&&~~~~\begin{pmatrix}
-ip_{z}-im-E\pm i\varepsilon & -ip_{x}+p_{y}\\
-ip_{x}-p_{y} & ip_{z}-im-E\pm i\varepsilon\\
\label{38}
\end{pmatrix}.
\end{eqnarray}
\end{small}
Thus, the Green's function is given as
\begin{small}
\begin{eqnarray}
&&~~~~G^{\pm}_{\varepsilon}(\vec{x},\vec{x}')=H_{0}^{-1}\delta(\vec{x}-\vec{x}')\nonumber\\
&&~~~~~~~~=H_{0}^{-1}\frac{1}{(2\pi)^{3}}\int d^{3} pe^{-i\vec{p}\cdot(\vec{x}-\vec{x}')}\nonumber\\
&&~~~~~~~~=\frac{1}{(2\pi)^{3}}\int d^{3}p\, \textit{diag}
(H_{01}^{-1},H_{02}^{-1},H_{01}^{-1},\nonumber\\
&&~~~~~~~~~~
H_{02}^{-1},H_{01}^{-1},H_{02}^{-1},H_{01}^{-1},H_{02}^{-1})\cdot e^{-i\vec{p}\cdot(\vec{x}-\vec{x}')},
\label{39}
\end{eqnarray}
\end{small}
and the scattering formula of Eq.\,(35) becomes
\begin{small}
\begin{eqnarray}
~~\Psi(\vec{x},t)&& \approx\varphi(\vec{x},t)-\frac{1}{(2\pi)^{3}}\lim\limits_{\varepsilon\rightarrow 0}\int d^{3}x'\int d^{3}p\nonumber\\
&&\textit{diag}(H_{01}^{-1},H_{02}^{-1},H_{01}^{-1},H_{02}^{-1},H_{01}^{-1},\nonumber\\
&& H_{02}^{-1},H_{01}^{-1},H_{02}^{-1})e^{-i\vec{p}\cdot(\vec{x}-\vec{x}')} V(\vec{x}')\varphi(\vec{x}',t).
\label{40}
\end{eqnarray}
\end{small}
where
$\varphi(\vec{x},t)=e^{iwt}\varphi_{0}(\vec{x})$ is the eigenfunction mentioned above. Letting $S\equiv
\begin{pmatrix}
1\\
0\\
\end{pmatrix}
 $,
we obtain
\begin{equation}
~~~~~~~~\varphi_{0}(\vec{x})=e^{ik_{z}z}S(1, 0, 1, 0, 1, 0, 1, 0)^{T}.\label{41}
\end{equation}
Next, we're going to derive
$ V(\vec{x})\varphi_{0}(\vec{x})\equiv P_{*}(\vec{x})\varphi_{0}(\vec{x})$.
It should be noticed that $P_{\ast}$ is a $[8\times8]$ matrix, while $\varphi_{0}(\vec{x})$ is a $[8\times 1]$ matrix.  If both of $S$ and $\sigma$ are assumed to be numbers, the matrix elements of $[P_{*}(\vec{x})\varphi_{0}(\vec{x})]$ would be given as follows.

1. For the first row,
\begin{small}
\begin{eqnarray}
~~~[P_{*}(\vec{x})\varphi_{0}(\vec{x})]_{1}&&=(\frac{\partial\Phi}{\partial x}+\frac{\partial\Phi}{\partial y}+\frac{\partial\Phi}{\partial z})e^{-ik_{z}z}S\nonumber\\
&&~~~=e^{-ik_{z}z}
\begin{pmatrix}
\frac{\partial\Phi}{\partial x}+\frac{\partial\Phi}{\partial y}+\frac{\partial\Phi}{\partial z}\\
0\\
\end{pmatrix},
\label{42}
\end{eqnarray}
\end{small}
2. For the second row,
\begin{small}
\begin{eqnarray}
~~~~~[P_{*}(\vec{x})\varphi_{0}(\vec{x})]_{2}&&=-3(\frac{\partial\Phi}{\partial x}\sigma_{1}+\frac{\partial\Phi}{\partial y}\sigma_{2}+\frac{\partial\Phi}{\partial z}\sigma_{3})\nonumber\\
&&\times e^{-ik_{z}z}S= -3e^{-ik_{z}z}
\begin{pmatrix}
\frac{\partial\Phi}{\partial z}\\
\frac{\partial\Phi}{\partial x}+i\frac{\partial\Phi}{\partial y}\\
\end{pmatrix},
\label{43}
\end{eqnarray}
\end{small}
3. For the third row
\begin{small}
\begin{eqnarray}
~~~~~~[ P_{*}(\vec{x})\varphi_{0}(\vec{x})]_{3}=\frac{\partial\Phi}{\partial x}e^{-ik_{z}z}S=e^{-ik_{z}z}
\begin{pmatrix}
\frac{\partial\Phi}{\partial x}\\
0\\
\end{pmatrix},
\label{44}
\end{eqnarray}
\end{small}
4. For the forth row
\begin{small}
\begin{eqnarray}
&&~~[ P_{*}(\vec{x})\varphi_{0}(\vec{x})]_{4}=[(3\frac{\partial\Phi}{\partial x}-\frac{\partial\Phi}{\partial y}-\frac{\partial\Phi}{\partial z})\sigma_{1}\nonumber\\
&&~~~~+(3\frac{\partial\Phi}{\partial y}+\frac{\partial\Phi}{\partial x})\sigma_{2}+(3\frac{\partial\Phi}{\partial z}+\frac{\partial\Phi}{\partial x})\sigma_{3}]e^{-ik_{z}z}S \nonumber\\
&&~~~~=e^{-ik_{z}z}
\begin{pmatrix}
3\frac{\partial\Phi}{\partial z}+\frac{\partial\Phi}{\partial x}\\
(3+i)\frac{\partial\Phi}{\partial x}+(-1+3i)\frac{\partial\Phi}{\partial y}-\frac{\partial\Phi}{\partial z}\\
\end{pmatrix},
\label{45}
\end{eqnarray}
\end{small}
5. For the fifth row,
\begin{small}
\begin{eqnarray}
~~~~~~~~[ P_{*}(\vec{x})\varphi_{0}(\vec{x})]_{5}=\frac{\partial\Phi}{\partial y}e^{-ik_{z}z}S=e^{-ik_{z}z}
\begin{pmatrix}
\frac{\partial\Phi}{\partial y}\\
0\\
\end{pmatrix},
\label{46}
\end{eqnarray}
\end{small}
6. For the sixth row
\begin{small}
\begin{eqnarray}
&&~~[ P_{*}(\vec{x})\varphi_{0}(\vec{x})]_{6}=[(3\frac{\partial\Phi}{\partial x}+\frac{\partial\Phi}{\partial y})\sigma_{1}+(3\frac{\partial\Phi}{\partial y}\nonumber\\
&&~~~~-\frac{\partial\Phi}{\partial x}-\frac{\partial\Phi}{\partial z})\sigma_{2}
+(3\frac{\partial\Phi}{\partial z}+\frac{\partial\Phi}{\partial y})\sigma_{3}]e^{-ik_{z}z}S \nonumber\\
&&~~~~= e^{-ik_{z}z}
\begin{pmatrix}
3\frac{\partial\Phi}{\partial z}+\frac{\partial\Phi}{\partial y}\\
(3-i)\frac{\partial\Phi}{\partial x}+(1+3i)\frac{\partial\Phi}{\partial y}-i\frac{\partial\Phi}{\partial z}\\
\end{pmatrix}
\label{47}
\end{eqnarray}
\end{small}
7. For the seventh row,
\begin{small}
\begin{eqnarray}
~~~~~~[P_{*}(\vec{x})\varphi_{0}(\vec{x})]_{7}=\frac{\partial\Phi}{\partial z}e^{-ik_{z}z}S=e^{-ik_{z}z}
\begin{pmatrix}
\frac{\partial\Phi}{\partial z}\\
0\\
\end{pmatrix},
\label{48}
\end{eqnarray}
\end{small}
8. For the eighth row,
\begin{small}
\begin{eqnarray}
&&~~~[ P_{*}(\vec{x})\varphi_{0}(\vec{x})]_{8}=[(3\frac{\partial\Phi}{\partial x}+\frac{\partial\Phi}{\partial z})\sigma_{1}+(3\frac{\partial\Phi}{\partial y}\nonumber\\
&&~~~~+\frac{\partial\Phi}{\partial z})\sigma_{2}
+(3\frac{\partial\Phi}{\partial z}-\frac{\partial\Phi}{\partial x}-\frac{\partial\Phi}{\partial y})\sigma_{3}]e^{-ik_{z}z}S \nonumber\\
&&~~~~= e^{-ik_{z}z}
\begin{pmatrix}
3\frac{\partial\Phi}{\partial z}-\frac{\partial\Phi}{\partial x}-\frac{\partial\Phi}{\partial y}\\
3\frac{\partial\Phi}{\partial x}+3i\frac{\partial\Phi}{\partial y}+(1+i)\frac{\partial\Phi}{\partial z}.\\
\end{pmatrix}
\label{49}
\end{eqnarray}
\end{small}
We write the scattering solution of the Dirac equation
\begin{small}
\begin{eqnarray}
 &&\Psi(\vec{x},t)\approx e^{i\omega t}\varphi_{0}(\vec{x})-e^{i\omega t}\lim\limits_{\varepsilon\rightarrow 0}\frac{1}{(2\pi)^{3}}\int d^{3}p \nonumber\\
&&~~~~\textit{diag}(H_{01}^{-1}, H_{02}^{-1},H_{01}^{-1},H_{02}^{-1},
H_{01}^{-1},H_{02}^{-1},\nonumber\\
&&~~~~ H_{01}^{-1},H_{02}^{-1})e^{-i\vec{p}\cdot\vec{x}}\int d^{3}\vec{x}'e^{i\vec{p}\cdot\vec{x}'}
P_{*}(\vec{x}') \varphi_{0}(\vec{x}').
\label{50}
\end{eqnarray}
\end{small}
Defining $M\equiv\int d^{3}\vec{x}'e^{i\vec{p}\cdot\vec{x}'}P_{*}(\vec{x}')\varphi_{0}(\vec{x}')$, which is a $[8\times1]$ matrix,
the eight matrix entries are given in Appendix B.

Let us continue exploring the matrix entries $M_{1}\sim M_{8}$. Firstly, the relations between Cartesian coordinates $x, y$, and $z$ and oblate elliptic coordinates $\xi, \eta$ and $\varphi$ are given as follows
\begin{eqnarray}
&&~~~~x=\rho_{0}\sqrt{(1+\xi^{2})(1-\eta^{2})}\cos \varphi,\,\nonumber\\
&&~~~~y=\rho_{0}\sqrt{(1+\xi^{2})(1-\eta^{2})}\sin\varphi,\,\nonumber\\
&&~~~~z=\rho_{0}\xi\eta,~~(0\leq\xi<\infty,~-1\leq\eta\leq1)
\label{51}
\end{eqnarray}
where $\rho_{0}$ is the focal length of a rotating spheroid, and the azimuth $\varphi\in(0, 2\pi)$.
By making a coordinate transformation $(x, y, z)\rightarrow(\xi,\eta,\varphi)$,
we get $d^{3}x\equiv dxdydz=\rho_{0}^{3}(\xi^{2}+\eta^{2})d\xi d\eta d\varphi$.
When $\xi>\xi_{0}$, the gravitational potential $\Phi$ of a rotating spheroid with
constant density $\mu$ is given by
\begin{small}
\begin{eqnarray}
~~\Phi=-\frac{M}{\rho_{0}}\{{\rm arccot}{\xi}+\frac{3}{4}[\xi-(\xi^{2}+\frac{1}{3}){\rm arccot}{\xi}](1-3\eta^{2})\},\label{52}
\end{eqnarray}
\end{small}
where $M$ is the mass of an ellipsoid star, which can be estimated as $M=4/3\pi
a^{2}c\mu=4/3\times\pi{\rho_{0}}^{3}\mu(1+{\xi_{0}}^{2})\xi_{0}$,($a=b=\rho_{0}\sqrt{1+\xi_{0}^{2}}$ is the semi-axis length in the equatorial plane, and $c=\rho_{0}\xi_{0}$ is the semi-axis length in the axis of rotation). When $0\leq\xi\leq\xi_{0}$, the gravitational potential $\Phi$ becomes
\begin{small}
\begin{eqnarray}
&&~~~~~\Phi=V_{0}+\frac{1}{2}\Omega^{2}\rho_{0}^{2}(1+\xi^{2})(1-\eta^{2})-C\nonumber\\
&&~~~~~~ \times \left[1-\frac{\rho_{0}^{2}
(1+\xi^{2})(1-\eta^{2})}{a^{2}}+C\frac{\rho_{0}^{2}\xi^{2}\eta^{2}}{c^{2}}\right],\label{53}
\end{eqnarray}
\end{small}
where $V_{0}=A(\xi_{0})$, $\Omega=\sqrt{2B(\xi_{0})}$ and
\begin{small}
\begin{eqnarray}
&& A=\frac{-3M[(\xi^{2}+1){\rm arc cot}\xi-\xi]}{2\rho_{0}},~B=\frac{3M\times}{4{\rho_{0}}^{3}(1+\xi^{2})}\nonumber\\
&& ~~ [(3\xi^{2}+1){\rm arc
cot}\xi-3\xi], C=\frac{3M\xi_{0}(1-\xi_{0} {\rm arc cot}\xi_{0})}{2\rho_{0}}.
\label{54}
\end{eqnarray}
\end{small}
For details about the gravitational potential  $\Phi$, see Chandrasekhar \cite{b32}.
From  Eq.\,(B6), we get the matrix entry
\begin{small}
\begin{eqnarray}
~~~~~M_{1}&&=-\int_{0}^{\infty}\int_{-1}^{1}\int_{0}^{2\pi}d\xi d\eta d\varphi\rho_{0}^{3}(\xi^{2}+\eta^{2})\times  \nonumber\\
&& ~~{\rm exp}[ip_{x}\rho_{0}\sqrt{(1+\xi^{2})(1-\eta^{2})}\cos\varphi+ \nonumber\\
&&~~ip_{y}\rho_{0}\sqrt{(1+\xi^{2})(1-\eta^{2})}\sin\varphi+i(p_{z}-k_{z})\rho_{0}\xi\eta]
\nonumber\\
&&~~\times\begin{pmatrix}
ip_{x}+ip_{y}+i(p_{z}-k_{z})\\
0\\
\end{pmatrix}
\Phi.
\label{55}
\end{eqnarray}
\end{small}
For convenience, we introduce the notation $T$ to denote the following integral,
\begin{small}
\begin{eqnarray}
~~~~~T&&=-\int_{0}^{\infty}\int_{-1}^{1}\int_{0}^{2\pi}d\xi d\eta d\varphi\rho_{0}^{3}(\xi^{2}+\eta^{2})\times  \nonumber\\
&&~~~~{\rm exp}[ip_{x}\rho_{0}\sqrt{(1+\xi^{2})(1-\eta^{2})}\sin\varphi+ip_{y}\rho_{0} \nonumber\\
&&~~~~\sqrt{(1+\xi^{2})(1-\eta^{2})}\cos\varphi+i(p_{z}-k_{z})\rho_{0}\xi\eta]
\Phi.
\label{56}
\end{eqnarray}
\end{small}
Thus, the matrix entry of $M_{1}$ becomes
\begin{small}
\begin{eqnarray}
~~~~~~~M_{1}&&\equiv
\begin{pmatrix}
ip_{x}+ip_{y}+i(p_{z}-k_{z})\\
0\\
\end{pmatrix}
T.
\label{57}
\end{eqnarray}
\end{small}
In the same way, we get the other seven matrix entries
\begin{small}
\begin{eqnarray}
~~~~~~~M_{2}&&\equiv
\begin{pmatrix}
-3i(p_{z}-k_{z})\\
-3ip_{x}+3p_{y}\\
\end{pmatrix}
T,~~~~M_{3}\equiv
\begin{pmatrix}
ip_{x}\\
0\\
\end{pmatrix}
T,
\label{58}
\end{eqnarray}
\end{small}
\begin{small}
\begin{eqnarray}
~~~~~~~M_{4}\equiv
\begin{pmatrix}
3i(p_{z}-k_{z})+ip_{x}\\
(3+i)ip_{x}+(-i+3i)ip_{y}-i(p_{z}-k_{z})\\
\end{pmatrix}
T,
\label{59}
\end{eqnarray}
\end{small}
\begin{small}
\begin{eqnarray}
~~~~~~~M_{5}\equiv
\begin{pmatrix}
ip_{y}\\
0\\
\end{pmatrix}
T,~~M_{7}\equiv
\begin{pmatrix}
i(p_{z}-k_{z})\\
0\\
\end{pmatrix}
T,
\label{60}
\end{eqnarray}
\end{small}
\begin{small}
\begin{eqnarray}
~~~~~~~~M_{6}\equiv
\begin{pmatrix}
3i(p_{z}-k_{z})+ip_{y}\\
(3-i)ip_{x}+(1+3i)ip_{y}+(p_{z}-k_{z})\\
\end{pmatrix}
T,
\label{61}
\end{eqnarray}
\end{small}
and
\begin{small}
\begin{eqnarray}
~~~~~~~~~M_{8}\equiv
\begin{pmatrix}
3i(p_{z}-k_{z})-ip_{x}-ip_{y}\\
3ip_{x}-3p_{y}+(1+i)i(p_{z}-k_{z})\\
\end{pmatrix}
T.
\label{62}
\end{eqnarray}
\end{small}
 Since the system we are studying is a rotating ellipsoid star, its gravitational potential $\Phi$ does not involve azimuth $\varphi$. Seeing from Eq.\,(55)-(62), the matrix entries of $M$ and the symbol $T$ are functions of $p_{x}$, $p_{y}$ and $p_{z}$. In the following, the scattering solution to the Dirac equation will be reexpressed with $M$ or $T$.  Then we have
\begin{small}
\begin{eqnarray}
 &&\Psi(\vec{x},t)=e^{i\omega t}\varphi_{0}(\vec{x})-e^{i\omega t}\lim\limits_{\varepsilon\rightarrow 0}\frac{1}{(2\pi)^{3}}\int d^{3}p \textit{diag}(H_{01}^{-1}, \nonumber\\
&&~ H_{02}^{-1}, H_{01}^{-1},H_{02}^{-1},H_{01}^{-1},H_{02}^{-1},H_{01}^{-1},H_{02}^{-1}) e^{-i\vec{p}\cdot\vec{x}}M.
\label{63}
\end{eqnarray}
\end{small}
where the expression of $\Psi(\vec{x},t)$$=(\phi^{(0)},\chi^{(0)},\phi^{(1)},\chi^{(1)},$\\
 $ \phi^{(2)},\chi^{(2)},\phi^{(3)},\chi^{(3)})$ is used.
\subsection{A special case: Maclaurin spheroids}
In this part, we will study the scattering of Dirac spinors in a special limiting case of Maclaurin spheroids. To understand the properties of Maclaurin spheroids, at first, let us discuss the motion of scalar particles in a rotating Maclaurin spheroid belonging to a special class of boson stars. When the distance between the center of the spheroid and a scalar particle, $r$, is larger than the stellar radius $R$, the exterior potential is given by $\Phi_{\rm ex}=-M/r$, corresponding to its partial derivative $(-\nabla \Phi_{\rm ext})_{r}=M/r^{2}$ and the acceleration of gravity, $a_{r}=-M/r^{2}$~\cite{b30}. From the Kepler's law, we get the angular velocity of the particle $\omega^{2}=M/r^{3}$\,(or $M/r^{2}=r\omega^{2}$ from the Newton Second Law of Motion). Similarly, when $(r\leq R$, the interior potential is given by $\Phi_{\rm in}=-2\pi\mu(R^{2}-r^{2}/3)$, corresponding to $(\nabla \Phi_{\rm in})_{r}=4/3\times\pi\mu r$ and $\omega^{2}=4/3 \times\pi\mu$.

According to statistical mechanics, the scattering cross section, also known as the collision section, is a physical quantity describing the scattering probability of microscopic particles. The dimension of the scattering cross-section is the same as that of the area. One core issue of scattering theory is to solve the scattering amplitude by studying the probability of the particles being scattered to the unit solid angle in the direction of $(p,\theta,\varphi)$. This probability can be expressed by the scattering differential cross-section $\sigma(p,\theta,\varphi)$, determined by the amplitude of the spherical scattered wave $ f(p,\theta,\varphi)$, i.e. $\sigma(p,\theta,\varphi)=|f(p,\theta,\varphi)|^{2}$.

In order to evaluate $T$ in a rotating Maclaurin spheroid, we define another new quantity as $\vec{p}_{0}=-p_{x}\vec{e}_{x}-p_{y}\vec{e}_{y}-(p_{z}-k_{z})\vec{e}_{z}$\, here
$\vec{e}_{x}$, $\vec{e}_{y}$ and $\vec{e}_{z}$ are the unit vectors in the $x$, $y$ and $z$ directions, respectively, then we have
\begin{small}
\begin{eqnarray}
~~~T&&=\int _{-\infty}^{+\infty}d^{3}\vec{x}e^{i\vec{p}\cdot\vec{x}-ip_{z}z}\Phi\nonumber\\
&&=\int_{0}^{R}dr\int_{0}^{\pi}d\theta\int_{0}^{2\pi}d\varphi r^{2}\sin \theta e^{-ip_{0}r\cos\theta} \nonumber\\
&&\times [-2\pi\mu(R^{2}-\frac{r^{2}}{3})]+\int_{R}^{+\infty}dr\int_{0}^{\pi}d\theta\int_{0}^{2\pi}d\varphi \nonumber\\
&&\times r^{2}\sin \theta e^{-ip_{0}r\cos\theta}(-\frac{M}{r})\nonumber\\
&&=\frac{8\pi^{2}\mu R}{P_{0}^{2}}(\frac{R^{2}}{3}+\frac{1}{p_{0}^{2}})\cos(p_{0}R)\nonumber\\
&&+ \frac{4\pi M}{p_{0}^{2}}(\cos\infty-\cos(p_{0}R))\nonumber\\
&&\approx \frac{8\pi^{2}\mu R}{P_{0}^{2}}(\frac{R^{2}}{3}+\frac{1}{p_{0}^{2}})\cos(p_{0}R)-\frac{4\pi M}{p_{0}^{2}}\cos(p_{0}R)\nonumber\\
&&=(-\frac{8\pi^{2}\mu R^{3}}{3p_{0}^{2}}+\frac{8\pi^{2}\mu R}{p_{0}^{4}})\cos(p_{0}R).
\label{64}
\end{eqnarray}
\end{small}
Since the scattering field is anisotropic, we can give the values of the scattering amplitude, $f_{i}(p,\theta,\varphi)$\,$(i=0, 1, 2, 3, 4, 5, 6, 7)$.
Here we consider a simple case of $i=0$, i.e. $\varphi^{(0)}$.
The scattering amplitude $f_{0}(p, \theta, \varphi)$ is determined by the following expression:
\begin{small}
\begin{eqnarray}
&& ~f_{0}(p,\theta,\varphi)=\lim\limits_{\varepsilon\rightarrow 0}\frac{1}{(ik_{z} \mp i\varepsilon)^{2}+p^{2}}\nonumber\\
&&~~~~ \times\begin{pmatrix}
-ip_{z}-ik_{z}\pm i\varepsilon & -ip_{x}+p_{y}\\
-ip_{x}-p_{y} & ip_{z}-ik_{z}\pm i\varepsilon\nonumber \\
\end{pmatrix} \nonumber \\
&& ~~~~\times\begin{pmatrix}
ip_{x}+ip_{y}+i(p_{z}-k_{z}) \\
0\\
\end{pmatrix}T\nonumber\\
&&~~= \lim \limits_{\varepsilon\rightarrow 0} \frac{1}{(ik_{z}{\brd \mp} i\varepsilon)^{2}+p^{2}}\times\nonumber\\
&&~~
\begin{pmatrix}
-ip\cos\theta-ik_{z}\pm i\varepsilon & -ip\cos\varphi\sin\theta+p\sin \varphi\sin\theta\\
-ip\cos\varphi\sin\theta-p\sin\varphi\sin\theta & ip\cos\theta-ik_{z}\pm i\varepsilon \\
\end{pmatrix}\nonumber\\
&& ~~~ \times
\begin{pmatrix}
ip\cos\varphi\sin\theta+ip\sin\varphi\sin\theta+i(p\cos\theta-k_{z})\\
0\\
\end{pmatrix}
T.
\label{65}
\end{eqnarray}
\end{small}
We also obtain the average value of $f_{0}(p, \theta, \varphi)$,
\begin{small}
\begin{eqnarray}
\bar{f}_{0}(p, \theta)=\frac{1}{2\pi}\int_{0}^{2\pi}f_{0}(p, \theta, \varphi)d\varphi=
\begin{pmatrix}
\cos^{2}\theta\\
\frac{1}{2}(1-i)\sin^{2}\theta\\
\end{pmatrix}
T.
\label{66}
\end{eqnarray}
\end{small}
The probability of scattering particle in the interval of $p\sim p+dp$ and $\theta\sim \theta+d\theta$ is determined by
\begin{eqnarray}
\quad \quad \quad \quad \frac{dW_{i}}{d^{3}p}=\frac{dW_{i}}{p^{2}\sin\theta dpd\theta}=
|\bar{f}_{i}(p,\theta)|^{2}.
\label{67}
\end{eqnarray}
Taking the long-wavelength approximation $k_{z}\rightarrow 0$, and paying attention to
$\vec{p}_{0}=-p_{x}\vec{e}_{x}-p_{y}\vec{e}_{y}-(p_{z}-k_{z})\vec{e}_{z}$,
we get the first component of $\bar{f}_{0}(p,\theta)$,
\begin{equation}
~~\bar{f}_{0}^{+}(p,\theta)=\cos^{2}\theta
[-\frac{8\pi^{2}\mu R^{3}}{3p^{2}}+
\frac{8 \pi^{2}\mu R}{p^{4}}]\cos(pR),
\label{68}
\end{equation}
and its corresponding scattering cross-section,
\begin{small}
\begin{eqnarray}
&&~~~~\bar{\sigma}_{0}^{+}(p,\theta)=|f_{0}^{+}(p,\theta)|^{2}\nonumber\\
&&~~~~~~~=\cos^{4}\theta
[-\frac{8 \pi^{2}\mu R^{3}}{3p^{2}}+
\frac{8 \pi^{2}\mu R}{p^{4}}]^{2}\cos^{2}(pR).
\label{69}
\end{eqnarray}
\end{small}
In the same way, the second component of the scattering amplitude is given as
\begin{small}
\begin{eqnarray}
\bar{f}_{0}^{-}(p,\theta)=\frac{1}{2}(1-i)\sin^{2}\theta[-\frac{8\pi^{2}\mu R^{3}}{3p^{2}}+
\frac{8\pi^{2}\mu R}{p^{4}}]\cos(pR),
\label{70}
\end{eqnarray}
\end{small}
corresponding to the scattering cross-section
\begin{small}
\begin{eqnarray}
~~~\bar{\sigma}_{0}^{-}(p,\theta)=\frac{1}{2}\sin^{4}\theta
[-\frac{8\pi^{2}\mu R^{3}}{3p^{2}}+
\frac{8\pi^{2}\mu R}{p^{4}}]^{2}\cos^{2}(pR).
\label{71}
\end{eqnarray}
\end{small}
Substituting Eq.\,(64) into the scattering solution, we find that the scattering cross-section is proportional to $\mu^{2}$, and depends on the radius of the rotating star. In addition, we can determine the constant density $\mu$ from observations, $\bar{\sigma}_{0}^{\pm}(p,\theta)$. From the above expression, it is obvious that the higher the density $\mu$, the larger the sensitivity of scattering amplitude $\bar{\sigma}_{0}^{\pm}(p,\theta)$ with regard to $\mu$. Notice that above we have chosen the long-wavelength approximation $E^{0}=-im\pm ik\approx -im$, and the star is not too compact. The properties of white dwarfs (WDs) in Einstein-$\wedge$ gravity were investigated by Liu and L\"{u}~\cite{b35}.  Considering the temperature effects~\cite{b36}, the massive WD, with a mass of about 0.61 $M_{\bigodot}$ and a radius of about 6000 km, has an average density of about $10^{6}$\,g~cm$^{-3}$, while the lower-mass WD, with a mass of about 0.21 $M_{\bigodot}$ and a radius of about 10,000 km, has an average density of about $10^{5}$\,g~cm$^{-3}$, where $M_{\bigodot}$ is the mass of the sun. Thus, our results are reasonable for WDs.

Similar to $\phi^{(0)}$, the scattering amplitudes for other seven quantities $\chi^{(0)}, \phi^{(1)}, \chi^{(1)}, \phi^{(2)}$, $\chi^{(2)},\phi^{(3)}$ and $\chi^{(3)}$ also can be obtained. However, due to limited space, the expressions of these seven scattering amplitudes will not be given in detail. Next, we shall continue to take the long-wavelength approximation. After a long but straightforward calculation, the scattering amplitudes are given as follows:
\begin{eqnarray}
&&~~\bar{f}_{0*}(p, \theta)=\frac{1}{2\pi}\int_{0}^{2\pi}f_{0*}(p, \theta, \varphi)d\varphi \nonumber\\
&&~~~~=\frac{1}{p^{2}-4m^{2}}
\begin{pmatrix}
p^{2}\cos^{2}\theta+2imp\cos\theta\\
\frac{1}{2}(1-i)p^{2}\sin^{2}\theta\\
\end{pmatrix}
T\nonumber\\
&&~~~\equiv\begin{pmatrix}
 \bar{f}_{0*}^{+}(p, \theta)\\
\bar{f}_{0*}^{-}(p, \theta)\\
\end{pmatrix}
\label{72}
\end{eqnarray}
 From the above expression, we obtain the two components of $\bar{\sigma}_{0*}(p,\theta)$,
\begin{eqnarray}
&&~~\bar{\sigma}_{0*}^{+}(p,\theta)=|\bar{f}_{0*}^{+}(p,\theta)|^{2}\nonumber\\
&&~~~~=\frac{1}{(p^{2}-4m^{2})^{2}}(p^{4}\cos^{4}
\theta+4m^{2}p^{2}\cos^{2}\theta)\nonumber\\
&&~~~~\times [-\frac{8\pi^{2}\mu R^{3}}{3p^{2}}+
\frac{8\pi^{2}\mu R}{p^{4}}]^{2}\cos^{2}(pR),
 \label{73}
\end{eqnarray}
and
\begin{eqnarray}
&&~~\bar{\sigma}_{0*}^{-}(p,\theta)=|\bar{f}_{0*}^{-}(p,\theta)|^{2}\nonumber\\
&&~~~~=\frac{1}{2(p^{2}-4m^{2})^{2}}\times p^{4}\sin^{4}
\theta\nonumber\\
&&~~~~\times [-\frac{8\pi^{2}\mu R^{3}}{3p^{2}}+
\frac{8\pi^{2}\mu R}{p^{4}}]^{2}\cos^{2}(pR).
 \label{74}
\end{eqnarray}
From the above, we find that scattering cross sections $\bar{\sigma}_{0}^{\pm}(E=im\pm ik\approx im)$ are independent of the mass of particles, $m$, while other scattering cross sections $\bar{\sigma}_{0*}^{\pm}(E=-im\pm ik\approx -im)$ depend on $m$. Therefore the observations of scattering cross sections can determine which form of energy ($E=\pm im\pm ik\approx \pm im$) should be adopted in our physical systems.
\section{Summary and outlook}
In this work, we have studied the scattering of spinors in the Dirac equation in detail, and discussed the scattering issue in the limit case of rotating Maclaurin spheroids, and established a relationship between the scattering cross section $\sigma$ and the density of matter $\mu$. We also found: the higher the density $\mu$, the higher the sensitivity of the scattering cross section $\sigma$ to the change of $\mu$. The results can be applied to all stars that can be treated with the Newtonian approximation approximately. In the future we'll determine the constant density of the star from the observed values of $\sigma_{0}^{+}(p,\theta,\varphi)$.

 Here we provide the prospect of the follow-up work, and study the direction continuing being advanced. First, we'll find out the other seven scattering cross sections $\bar{\sigma}_{i}^{\pm}(p,\theta)(E^{0}=im\pm ik\approx im)$, and $\bar{\sigma}_{i^{*}}^{\pm}(p,\theta)(E^{0}=-im\pm ik\approx -im)$, $(i=1, 2, 3, 4, 5, 6, 7)$, thus the average density of the star will be determined; Secondly, we'll consider a more complex model on rotating spheroids with electromagnetic field. The scattering cross sections with and without electromagnetic(EM) fields maybe different. Thirdly, we will investigate the motion of particles inside an accretion disk around a rotating spheroid. Due to the gravity of a rotating spheroid, the disk might be possible. Finally, it is expected that we study the physical effects of a compact rotating spheroid, especially to solve the scattering solutions to the geodesic equation,scalar-field equation and spinor-field equation, respectively.For the expansion of the Newtonian analysis, we will be concerned with a metric perturbation $h_{\nu\upsilon}$ away from flat space-time, defined as $g_{\mu\nu}=\eta_{\mu\nu}+h_{\mu\nu}$, and explore the gravitational properties of rotating spheroids, as well as the scattering solution under supergravity.
\begin{acknowledgements}
We are grateful the referee for helpful comments. This work was supported by National Key Research and Development Program of China under grant number 2018YFA0404602, and Chinese National Science Foundation through grants No. 11673056 and 11173042, and Xinjiang Natural Science Foundation No.2018D01A24 and Guizhou provincial Science and Technology Project ([2017]7349, [2019]1241). This work was also supported by CAS Light of West China Program (No.2016-QNXZ-B-25), Xiaofeng Yang¡¯s Xinjiang Tianchi Bairen project and CAS Pioneer Hundred Talents Program.
\end{acknowledgements}

%


\begin{thebibliography}{}
%
%
\bibitem{b1}G. Walter, Wave Equation, 1990, Relativistic Quantum Mechanics
   (Springer, Berlin, 1990)
\bibitem{b2} A. Zee, Quantum Field Theory in a Nutshell, 2nd edn.
   (Princeton University Press, Princeton, 2010)
\bibitem{b3}L. E. Parker and D. J. Toms, Quantum Field theory in Curved space-time: Quantum  Fields and Gravity, P216-257, Cambridge University Press 2009; arXiv: gr-qc/0008033v1 (Appendix A).
\bibitem{b4}V. D. Dzhunushaliev, Spherically Symmetric Solution for Torsion and the Dirac equation in 5D spacetime, 1998, Int. J. Mod. Phys. D, 7, 909-915(1998)
\bibitem{b5}M. alimohammadi and A. Shariati, Neutrino oscillation in a space-time with torsion, 1999, Mod. Phys. Lett. A, 14, 267(1999)
\bibitem{b6}Z. Daniel, Freedman and Antoine van Proeyen, Supergravity (Cambridge University Press, Cambridge, 2012)
\bibitem{b7}P. van Nieuwenhuizen, Supergravity, Physics reports (Review Section of physics letters.)68(4),189-398 (1981)
\bibitem{b8}M. Alimohammadi and  A. Shariati;  Neutrino oscillation in a space-time with torsion. 1999, Mod.Phys.Lett. A. 14, 267-274(1999)
\bibitem{b9}V. M. Villalba, Dirac spinor in a nonstationary Godel-type cosmological Universe. 1993,  Mod. Phys. Lett. A, 8, 3011-3018(1993), arXiv: gr-qc/9309019
\bibitem{b10} S. Dolan; C. Doran and  A. Lasenby; Fermion scattering by a Schwarzschild black hole, 2006, Phys. Rev. D, 74, 064005(2006)
\bibitem{b11}S. De Leo and  P. Rotelli; Potential Scattering in Dirac Field Theory, 2009,  Eur.Phys. J. C, 62, 792-797(2009)
\bibitem{b12}V. Stefano; F. Luca and  C. Roberto; Dirac spinors in Bianchi-I f(R)-cosmology with torsion, 2011, J. Math. Phys. 52, 112502(2011)
\bibitem{b13}C. J. Bord\'{e}; J.-C. Houard and  A. Karasiewicz; Relativistic phase shifts for Diracparticles interacting with weak gravitational fields in matter-wave interferometers. Lect. Notes Phys. 562, 403-438(2011)
\bibitem{b14}N. Poplawski; Nonsingular, big-bounce cosmology from spinor-torsion coupling. (2012), Phys. Rev. D, 85, 107502, arXiv: gr-qc/1111.4595
\bibitem{b15} T. Daude and  N. Kamran; Local energy decay of massive Dirac fields in the 5D Myers-Perry metric, (2012), Classical and Quantum Gravity, 29, 14
\bibitem{b16} Y. Brihaye; T. Delsate;  N. Sawado and  H. Yoshii; Dirac equation for sphercially symmetric AdS5 space-time and application to a boson star in EGB gravity. (2014), arXiv:gr-qc/1410.7539
\bibitem{b17} V. E. Ambrus and  E. Winstanley; Dirac fermions on an anti-de Sitter background.AIP Conf. Proc. 1634, 40(2014)
\bibitem{b18}D. Bini; E. Bittencourt and  A. Geralico; Massless Dirac particles in the vacuum C-metric. Classical and Quantum Gravity, 32, 21,(2015) arXiv:gr-qc/1509.04878
\bibitem{b19} C. R\"{o}ken; The Massive Dirac Equation in the Kerr Geometry: Separability in Eddington-Finkelstein-type Coordinates and Asymptotics. General Relativity and Gravitation, 49, 39.(2017)  arXiv:gr-qc/1506.08038
\bibitem{b20}V. Dzhunushaliev and V. Folomeev; Dirac star in the presence of Maxwell and Proca fields.  Phys. Rev. D, 99, 104066(2019)
\bibitem{b21}R. R. S. Oliveira; Noninertial and spin effects on the 2D Dirac oscillator in the magnetic cosmic string background, (2019) arXiv:1906.07369
\bibitem{b22}D. Batic; M. Nowakowsk and  K. Morgan; The problem of embedded eigenvalues for the Dirac equation in the Schwarzschild black hole metric, (2016). Universe, 2, 31 arXiv:1701.03889v1
\bibitem{b23}G. V. Kraniotis; The massive Dirac equation in the Kerr-Newman-de Sitter and Kerr-Newman black hole spacetimes. J. Phys. Commun. 3, 035026,(2019) arXiv: 1801.03157
\bibitem{b24}B.-S. Jose Luis and K. Christian; Solutions of the massive Dirac equation in the near-horizon metric of the extremal five dimensional Myers-Perry black hole with equal angular momenta. Phys. Rev. D, 99, 024026, (2019)arXiv:1808.00503
\bibitem{b25}M.-A. Dariescu; C. Dariescu and C. Stelea; Heun-type solutions of the Klein-Gordon and Dirac equations in the Garnkle-Horowitz-Strominger dilaton black hole background, (2018) arXiv: 1812.06852
\bibitem{b26}Z.-W. Feng; Q.-C. Ding and S.-Z. Yang; Modified fermion tunneling from higher-dimensional charged AdS black hole in massive gravity. Eur. Phys. J. C, 79, 7(2019)
\bibitem{b27}K. Lin and S.-Z. Yang; Quantum tunnelling in charged black holes beyond the semi-classical approximation.  Europhys. Lett. 86, 20006(2019)
\bibitem{b28}S. Z. Yang and K. Lin; Modified fermiions tunnelling radiation from Kerr-Newman-de Sitter black hole(in chinese). Sci Sin-Phys Mech Astron,49, 019503, (2019) doi:10.1360/SSPMA2018-00307
\bibitem{b29}Z. F. Gao; D. L. Song; X.-D. Li; H. Shan and N. Wang,  The equilibrium equations of Boson-Fermi systems in the Newtonian approximation,  Astron. Nachr. 340, 241(2019)
\bibitem{b30} A.-B. Ahmad; I. Sakall; S. Kanzi; Solution of Dirac equation and greybody radiation around a regular Bardeen black hole surrounded by quintessence; Annals of Physics, Vol. 412 168026 (2020)
\bibitem{b31} B. Z. LLiev;  Fibre bundle formulation of nonrelativistic quantum mechanics, III Pictures adn integrals of motion. J. Phys. A. 34, 4935,(2001)arXiv: 9806046
\bibitem{b32}S. Chandrasekhar, ''Ellipsoidal Figures of Equilibrium'', (Yale University Press 1969)
\bibitem{b33}J. J. Sakurai, 2006, "Moden Quantum Mechanics" (Revised Edition),( World Publishing Corporation,New York,2006) ISBN:9787506273145
\bibitem{b34}V. G. Barlette; M.M. Leite and S.K. Adhikaris; Integral Equations of Scattering in One Dimention, Amm. J. Phys, 69(9), 1010(2001)
\bibitem{b35}H. L. Liu and G.-L. L\"{u}; Properties of white dwarfs in Einstein-$\wedge$ gravity, 2019,J. Cosmol. Astropart.Phys, 2019, 040(2019)
\bibitem{b36}G.-L. L\"{u}; C.-H. Zhu; Z. Wang; H. Liu; L. Li; D. Xie and J. Liu; Possible Formation Scenarios of ZTF J153932.16+502738.8¡ªA Gravitational Source Close to the Peak of LISA¡¯s Sensitivity. 2020, The Astrophysical Journal, 890, 69(2020)
\end{thebibliography}


\appendix{ }
\section{The concrete expressions of the Dirac equation in the weak gravitational potential }
 Here $i\gamma^{(c)}e_{(c)}^{\mu}D_{\mu}\Psi^{(b)}-m\Psi^{(b)}=0~(b=0,1,2,3)$ are written as follows
\begin{small}
\begin{eqnarray}
Q_{b}(\phi,\chi)
\begin{pmatrix}
1\\
0
\end{pmatrix}
+Q_{b}(\phi,\chi:\phi\leftrightarrow\chi)
\begin{pmatrix}
0\\
-1
\end{pmatrix}
-m
\begin{pmatrix}
\Phi^{(b)}\\
\chi^{(b)}
\end{pmatrix}
=0,\nonumber\\
\label{A1}
\end{eqnarray}
\end{small}
where
\begin{small}
\begin{eqnarray}
&&Q_{0}=i[\partial_{0}\phi^{(0)}+\frac{\partial\Phi}{\partial x}\phi^{(1)}+\frac{\partial\Phi}{\partial y}\phi^{(2)}+\frac{\partial\Phi}{\partial z}\phi^{(3)}]\nonumber\\
&&~~~+i\sigma_{1}[\partial_{1}\chi^{(0)}-\frac{\partial\Phi}{\partial x}\chi^{(1)}-\frac{\partial\Phi}{\partial x}\chi^{(2)}-\frac{\partial\Phi}{\partial x}\chi^{(3)}]\nonumber\\
&&~~~+i\sigma_{2}[\partial_{2}\chi^{(0)}-\frac{\partial\Phi}{\partial y}\chi^{(1)}-\frac{\partial\Phi}{\partial y}\chi^{(2)}-\frac{\partial\Phi}{\partial y}\chi^{(3)}]\nonumber\\
&&~~~+i\sigma_{3}[\partial_{3}\chi^{(0)}-\frac{\partial\Phi}{\partial z}\chi^{(1)}-\frac{\partial\Phi}{\partial z}\chi^{(2)}-\frac{\partial\Phi}{\partial z}\chi^{(3)}],
\label{A2}
\end{eqnarray}
\end{small}
\begin{small}
\begin{eqnarray}
&&Q_{1}=i[\partial_{0}\phi^{(1)}+\frac{\partial\Phi}{\partial x}\phi^{(0)}]+i\sigma_{1}[\partial_{1}\chi^{(1)}+
\nonumber\\
&&~~\frac{\partial\Phi}{\partial x}\chi^{(0)}+(\frac{\partial\Phi}{\partial x}-\frac{\partial\Phi}{\partial y})\chi^{(2)}+(\frac{\partial\Phi}{\partial x}-\frac{\partial\Phi}{\partial z})\chi^{(3)}]+\nonumber\\
&&~~i\sigma_{2}[\partial_{2}\chi^{(1)}+\frac{\partial\Phi}{\partial y}\chi^{(0)}+(\frac{\partial\Phi}{\partial y}+\frac{\partial\Phi}{\partial x})\chi^{(2)}+\frac{\partial\Phi}{\partial y}\chi^{(3)}]+\nonumber\\
&&~~i\sigma_{3}[\partial_{3}\chi^{(1)}+\frac{\partial\Phi}{\partial z}\chi^{(0)}+
\frac{\partial\Phi}{\partial z}\chi^{(2)}+(\frac{\partial\Phi}{\partial z}+\frac{\partial\Phi}{\partial x})\chi^{(3)}],
\label{A3}
\end{eqnarray}
\end{small}
\begin{small}
\begin{eqnarray}
&&Q_{2}=i[\partial_{0}\phi^{(2)}+\frac{\partial\Phi}{\partial y}\phi^{(0)}]+i\sigma_{1}\nonumber\\
&&~[\partial_{1}\chi^{(2)}+\frac{\partial\Phi}{\partial x}\chi^{(0)}+(\frac{\partial\Phi}{\partial x}+\frac{\partial\Phi}{\partial y})\chi^{(1)}+\frac{\partial\Phi}{\partial x}\chi^{(3)}]+i\sigma_{2}\nonumber\\
&&~[\partial_{2}\chi^{(2)}+\frac{\partial\Phi}{\partial y}\chi^{(0)}+(\frac{\partial\Phi}{\partial y}-\frac{\partial\Phi}{\partial x})\chi^{(1)}+(\frac{\partial\Phi}{\partial y}-\frac{\partial\Phi}{\partial z})\chi^{(3)}]\nonumber\\
&&+i\sigma_{3}[\partial_{3}\chi^{(2)}+\frac{\partial\Phi}{\partial z}\chi^{(0)}+\frac{\partial\Phi}{\partial z}\chi^{(1)}+(\frac{\partial\Phi}{\partial z}+\frac{\partial\Phi}{\partial y})\chi^{(3)}],
\label{A4}
\end{eqnarray}
\end{small}
and
\begin{small}
\begin{eqnarray}
&&Q_{3}=i[\partial_{0}\phi^{(3)}+\frac{\partial\Phi}{\partial z}\phi^{(0)}]+i\sigma_{1}\nonumber\\
&&~~[\partial_{1}\chi^{(3)}+\frac{\partial\Phi}{\partial x}\chi^{(0)}+(\frac{\partial\Phi}{\partial x}+\frac{\partial\Phi}{\partial z})\chi^{(1)}+\frac{\partial\Phi}{\partial x}\chi^{(2)}]\nonumber\\
&&~+i\sigma_{2}[\partial_{2}\chi^{(3)}+\frac{\partial\Phi}{\partial y}\chi^{(0)}+\frac{\partial\Phi}{\partial y}\chi^{(1)}+(\frac{\partial\Phi}{\partial x}+\frac{\partial\Phi}{\partial z})\chi^{(2)}]\nonumber\\
&&~+i\sigma_{3}[\partial_{3}\chi^{(3)}+\frac{\partial\Phi}{\partial z}\chi^{(0)}+(\frac{\partial\Phi}{\partial z}-\frac{\partial\Phi}{\partial x})\chi^{(1)}+(\frac{\partial\Phi}{\partial z}-\frac{\partial\Phi}{\partial y})\chi^{(2)}].\nonumber\\
\label{A5}
\end{eqnarray}
\end{small}

\section{ The matrix $M$ and the scattering solutions to the Dirac equation}
The eight matric entries are given as following:
\begin{small}
\begin{eqnarray}
~~~M_{1}&&\equiv\int d^{3}\vec{x}'e^{i\vec{p}\cdot\vec{x}'}[P_{*}(\vec{x}')\varphi_{0}(\vec{x}')]_{1}\nonumber\\
&&=\int d^{3}\vec{x}'e^{i\vec{p}\cdot\vec{x}'}e^{-ik_{z}z'}
\begin{pmatrix}
\frac{\partial\Phi}{\partial x'}+\frac{\partial\Phi}{\partial y'}+\frac{\partial\Phi}{\partial z'}\\
0\\
\end{pmatrix}\nonumber\\
&&=-\int d^{3}\vec{x}'e^{ip_{x}x'+ip_{y}y'+i(p_{z}-k_{z})z'}\nonumber\\
&&~~\times \begin{pmatrix}
ip_{x}+ip_{y}+i(p_{z}-k_{z})\\
0
\end{pmatrix}
\Phi,
\label{B1}
\end{eqnarray}
\end{small}
\begin{small}
\begin{eqnarray}
~~~M_{2}&&\equiv\int d^{3}\vec{x}'e^{i\vec{p}\cdot\vec{x}'}[P_{*}(\vec{x}')\varphi_{0}(\vec{x}')]_{2}\nonumber\\
&&=\int d^{3}\vec{x}'e^{i\vec{p}\cdot\vec{x}'}e^{-ik_{z}z'}
\begin{pmatrix}
3\frac{\partial\Phi}{\partial z'}\\
3\frac{\partial\Phi}{\partial x'}+3i\frac{\partial\Phi}{\partial y'}\\
\end{pmatrix} \nonumber \\
&&=-\int d^{3}\vec{x}'e^{ip_{x}x'+ip_{y}y'+i(p_{z}-k_{z})z'}\nonumber\\
&&\times \begin{pmatrix}
3i(p_{z}-k_{z})\\
3ip_{x}+3p_{y}
\end{pmatrix}
\Phi,\nonumber \\
\label{B2}
\end{eqnarray}
\end{small}
\begin{small}
\begin{eqnarray}
~~M_{3}&&\equiv\int d^{3}\vec{x}'e^{i\vec{p}\cdot\vec{x}'}[P_{*}(\vec{x}')\varphi_{0}(\vec{x}')]_{3}\nonumber\\
&&=\int d^{3}\vec{x}'e^{i\vec{p}\cdot\vec{x}'}e^{-ik_{z}z'}
\begin{pmatrix}
\frac{\partial\Phi}{\partial x'}\\
0\\
\end{pmatrix} \nonumber\\
&&~=-\int d^{3}\vec{x}'e^{ip_{x}x'+ip_{y}y'+i(p_{z}-k_{z})z'}
\begin{pmatrix}
ip_{x}\\
0\\
\end{pmatrix}
\Phi,
\label{B3}
\end{eqnarray}
\end{small}
\begin{small}
\begin{eqnarray}
~M_{4}&&\equiv\int d^{3}\vec{x}'e^{i\vec{p}\cdot\vec{x}'}[P_{*}(\vec{x}')\varphi_{0}(\vec{x}')]_{4}\nonumber\\
&&=\int d^{3}\vec{x}'e^{i\vec{p}\cdot\vec{x}'}e^{-ik_{z}z'}\nonumber\\
&&\times \begin{pmatrix}
3\frac{\partial\Phi}{\partial z'}+\frac{\partial\Phi}{\partial x'}\\
(3+i)\frac{\partial\Phi}{\partial x'}+(-1+3i)\frac{\partial\Phi}{\partial y'}-\frac{\partial \Phi}{\partial z'}\\
\end{pmatrix}\nonumber\\
&&=-\int d^{3}\vec{x}'e^{ip_{x}x'+ip_{y}y'+i(p_{z}-k_{z})z'}\nonumber\\
&&\times\small\begin{pmatrix}
3i(p_{z}-k_{z})+ip_{x}\\
(3+i)ip_{x}+(-1+3i)ip_{y}-i(p_{z}+k_{z})
\end{pmatrix}
\Phi,
\label{B4}
\end{eqnarray}
\end{small}
\begin{small}
\begin{eqnarray}
~~~~M_{5}&&\equiv\int d^{3}\vec{x}'e^{i\vec{p}\cdot\vec{x}'}[P_{*}(\vec{x}')\varphi_{0}(\vec{x}')]_{5}\nonumber\\
&&=\int d^{3}\vec{x}'e^{i\vec{p}\cdot\vec{x}'}e^{-ik_{z}z'}
\begin{pmatrix}
\frac{\partial\Phi}{\partial y'}\\
0\\
\end{pmatrix} \nonumber\\
&&=-\int d^{3}\vec{x}'e^{ip_{x}x'+ip_{y}y'+i(p_{z}-k_{z})z'}
\begin{pmatrix}
ip_{y}\\
0\\
\end{pmatrix}
\Phi,
\label{B5}
\end{eqnarray}
\end{small}
\begin{small}
\begin{eqnarray}
~~~~M_{6}&&\equiv\int d^{3}\vec{x}'e^{i\vec{p}\cdot\vec{x}'}[P_{*}(\vec{x}')\varphi_{0}(\vec{x}')]_{6}\nonumber\\
&&=\int d^{3}\vec{x}'e^{i\vec{p}\cdot\vec{x}'}e^{-ik_{z}z'}\nonumber\\
&&\times \begin{pmatrix}
3\frac{\partial\Phi}{\partial z'}+\frac{\partial\Phi}{\partial y'}\\
(3-i)\frac{\partial\Phi}{\partial x'}+(1+3i)\frac{\partial\Phi}{\partial y'}-i\frac{\partial \Phi}{\partial z'}\\
\end{pmatrix}\nonumber\\
&&=-\int d^{3}\vec{x}'e^{ip_{x}x'+ip_{y}y'+i(p_{z}-k_{z})z'}\nonumber\\
&&\times \small\begin{pmatrix}
3i(p_{z}-k_{z})+ip_{y}\\
(3-i)ip_{x}+(1+3i)ip_{y}{\brd +}i(p_{z}-k_{z})
\end{pmatrix}
\Phi,
\label{B6}
\end{eqnarray}
\end{small}
\begin{small}
\begin{eqnarray}
~~~~M_{7}&&\equiv\int d^{3}\vec{x}'e^{i\vec{p}\cdot\vec{x}'}[P_{*}(\vec{x}')\varphi_{0}(\vec{x}')]_{7}\nonumber\\
&&=\int d^{3}\vec{x}'e^{i\vec{p}\cdot\vec{x}'}e^{-ik_{z}z'}
\begin{pmatrix}
\frac{\partial\Phi}{\partial z'}\\
0\\
\end{pmatrix}\nonumber\\
&&=-\int d^{3}\vec{x}'e^{ip_{x}x'+ip_{y}y'+i(p_{z}-k_{z})z'}
 \begin{pmatrix}
i(p_{z}-k_{z})\\
0\\
\end{pmatrix}
\Phi,\nonumber\\
\label{B7}
\end{eqnarray}
\end{small}
and
\begin{small}
\begin{eqnarray}
~~~~M_{8}&&\equiv\int d^{3}\vec{x}'e^{i\vec{p}\cdot\vec{x}'}[P_{*}(\vec{x}')\varphi_{0}(\vec{x}')]_{8}\nonumber\\
&&=\int d^{3}\vec{x}'e^{i\vec{p}\cdot\vec{x}'}e^{-ik_{z}z'}\nonumber\\
&& \times\begin{pmatrix}
3\frac{\partial\Phi}{\partial z'}-\frac{\partial\Phi}{\partial x'}-\frac{\partial\Phi}{\partial y'}\\
3\frac{\partial\Phi}{\partial x'}+3i\frac{\partial\Phi}{\partial y'}+
(1+i)\frac{\partial \Phi}{\partial z'}\\
\end{pmatrix} \nonumber\\
&&=-\int d^{3}\vec{x}'e^{ip_{x}x'+ip_{y}y'+i(p_{z}-k_{z})z'}\nonumber\\
&&\times \small\begin{pmatrix}
3i(p_{z}-k_{z})-ip_{x}-ip_{y}\\
3ip_{x}-3p_{y}+(1+i)i(p_{z}-k_{z})\\
\end{pmatrix}
\Phi.
\label{B8}
\end{eqnarray}\\
\end{small}

The scattering solutions to the Dirac equation are given as
\begin{small}
\begin{eqnarray}
~&&\phi^{(0)}\approx e^{i\omega t-ik_{z}z}
\begin{pmatrix}
1\\
0\\
\end{pmatrix}\nonumber\\
&&~~~-e^{i\omega t}\lim\limits_{\varepsilon\rightarrow0}\frac{1}{(2\pi)^{3}}\int d^{3}pe^{-i\vec{p}\cdot\vec{x}}H_{01}^{-1}M_{1} \nonumber\\
&&~~~~=e^{i\omega t-ik_{z}z}
\begin{pmatrix}
1\\
0\\
\end{pmatrix}-e^{i\omega t}\lim\limits_{\varepsilon\rightarrow0}\frac{1}{(2\pi)^{3}}\int d^{3}pe^{-i\vec{p}\cdot\vec{x}}\nonumber\\
&&~~~~\frac{1}{(ik_{z}\mp i\varepsilon)^{2}+p^{2}}
\begin{pmatrix}
-ip_{z}-ik_{z}\pm i\varepsilon & -ip_{x}+p_{y}\\
-ip_{x}-p_{y} & ip_{z}-ik_{z}\pm i\varepsilon\\
\end{pmatrix}\nonumber\\
&&~~~~\times
\begin{pmatrix}
ip_{x}+ip_{y}+i(p_{z}-k_{z})\\
0\\
\end{pmatrix}
T,
\label{B9}
\end{eqnarray}
\end{small}
\begin{small}
\begin{eqnarray}
~~\chi^{(0)}&&\approx -e^{i\omega t}\lim\limits_{\varepsilon\rightarrow0}\frac{1}{(2\pi)^{3}}\int d^{3}pe^{-i\vec{p}\cdot\vec{x}}H_{02}^{-1}M_{2}\nonumber\\
&&=-e^{i\omega t}\lim\limits_{\varepsilon\rightarrow0}\frac{1}{(2\pi)^{3}}\int d^{3}pe^{-i\vec{p}\cdot\vec{x}}\nonumber\\
&&\times\frac{1}{(-2im-ik_{z}\pm i\varepsilon)^{2}+p^{2}}\times
\small\nonumber\\
&&\begin{pmatrix}
-ip_{z}-2im-ik_{z}\pm i\varepsilon & -ip_{x}+p_{y}\\
-ip_{x}-p_{y} & ip_{z}-2im-ik_{z}\pm i\varepsilon\\
\end{pmatrix}\nonumber\\
&&~\times
\begin{pmatrix}
-3i(p_{z}-k_{z})\\
-3ip_{x}+3p_{y}\\
\end{pmatrix}
T,
\label{B10}
\end{eqnarray}
\end{small}
\begin{small}
\begin{eqnarray}
~~\phi^{(1)}&&\approx e^{i\omega t-ik_{z}z}
\begin{pmatrix}
1\\
0\\
\end{pmatrix}\nonumber\\
&&-e^{i\omega t}\lim\limits_{\varepsilon\rightarrow0}\frac{1}{(2\pi)^{3}}\int d^{3}pe^{-i\vec{p}\cdot\vec{x}}H_{01}^{-1}M_{3}\nonumber\\
&&=e^{i\omega t-ik_{z}z}
\begin{pmatrix}
1\\
0\\
\end{pmatrix}
-e^{i\omega t}\lim\limits_{\varepsilon\rightarrow0}\frac{1}{(2\pi)^{3}}\nonumber\\
&&\times\int d^{3}pe^{-i\vec{p}\cdot\vec{x}}
\frac{1}{(ik_{z}\mp i\varepsilon)^{2}+p^{2}}\times \nonumber\\
&&\begin{pmatrix}
-ip_{z}-ik_{z}\pm i\varepsilon & -ip_{x}+p_{y}\\
-ip_{x}-p_{y} & ip_{z}-ik_{z}\pm i\varepsilon\\
\end{pmatrix}
\begin{pmatrix}
ip_{x}\\
0\\
\end{pmatrix}
T,\nonumber\\
\label{B11}
\end{eqnarray}
\end{small}
\begin{small}
\begin{eqnarray}
&&~~\chi^{(1)}\approx -e^{i\omega t}\lim\limits_{\varepsilon\rightarrow0}\frac{1}{(2\pi)^{3}}\int d^{3}pe^{-i\vec{p}\cdot\vec{x}}H_{02}^{-1}M_{4}\nonumber\\
&&~~~~~=-e^{i\omega t}\lim\limits_{\varepsilon\rightarrow0}\frac{1}{(2\pi)^{3}}\int d^{3}pe^{-i\vec{p}\cdot\vec{x}}\nonumber\\
&&~~~~~\times\frac{1}{(-2im-ik_{z}\pm i\varepsilon)^{2}+p^{2}}\nonumber\\
&&~~~~~\times\small\begin{pmatrix}
-ip_{z}-2im-ik_{z}\pm i\varepsilon & -ip_{x}+p_{y}\\
-ip_{x}-p_{y} & ip_{z}-2im-ik_{z}\pm i\varepsilon\\
\end{pmatrix}\nonumber\\
&&~~~~~~\times
\begin{pmatrix}
3i(p_{z}-k_{z})+ip_{x}\\
(3+i)ip_{x}+(-1+3i)ip_{y}-i(p_{z}-k_{z})\\
\end{pmatrix}
T,
\label{B12}
\end{eqnarray}
\end{small}
\begin{small}
\begin{eqnarray}
&&~~\phi^{(2)}\approx e^{i\omega t-ik_{z}z}
\begin{pmatrix}
1\\
0\\
\end{pmatrix}\nonumber\\
&&~~~~-e^{i\omega t}\lim\limits_{\varepsilon\rightarrow0}\frac{1}{(2\pi)^{3}}\int d^{3}pe^{-i\vec{p}\cdot\vec{x}}\frac{1}{(ik_{z}\mp i\varepsilon)^{2}+p^{2}}\nonumber\\
&&~~~~\times\begin{pmatrix}
-ip_{z}-ik_{z}\pm i\varepsilon & -ip_{x}+p_{y}\\
-ip_{x}-p_{y} & ip_{z}-ik_{z}\pm i\varepsilon\\
\end{pmatrix}
\begin{pmatrix}
ip_{y}\\
0\\
\end{pmatrix}
T,
\label{B13}
\end{eqnarray}
\end{small}
\begin{small}
\begin{eqnarray}
~~~\chi^{(2)}&&\approx -e^{i\omega t}\lim\limits_{\varepsilon\rightarrow0}\frac{1}{(2\pi)^{3}}\int d^{3}pe^{-i\vec{p}\cdot\vec{x}}H_{02}^{-1}M_{6}\nonumber\\
&&=-e^{i\omega t}\lim\limits_{\varepsilon\rightarrow0}\frac{1}{(2\pi)^{3}}\int d^{3}pe^{-i\vec{p}\cdot\vec{x}}\nonumber\\
&&\times\frac{1}{(-2im-ik_{z}\pm i\varepsilon)^{2}+p^{2}}\nonumber\\
&&~~\times\small\begin{pmatrix}
-ip_{z}-2im-ik_{z}\pm i\varepsilon & -ip_{x}+p_{y}\\
-ip_{x}-p_{y} & ip_{z}-2im-ik_{z}\pm i\varepsilon\\
\end{pmatrix}\nonumber\\
&&~~\times
\begin{pmatrix}
3i(p_{z}-k_{z})+ip_{y}\\
(3-i)ip_{x}+(1+3i)ip_{y}+(p_{z}-k_{z})\\
\end{pmatrix}
T,
\label{B14}
\end{eqnarray}
\end{small}
\begin{small}
\begin{eqnarray}
~~\phi^{(3)}&&\approx e^{i\omega t-ik_{z}z}
\begin{pmatrix}
1\\
0\\
\end{pmatrix}\nonumber\\
&&-e^{i\omega t}\lim\limits_{\varepsilon\rightarrow0}\frac{1}{(2\pi)^{3}}\int d^{3}pe^{-i\vec{p}\cdot\vec{x}}H_{01}^{-1}M_{7}\nonumber\\
&&~=e^{i\omega t-ik_{z}z}
\begin{pmatrix}
1\\
0\\
\end{pmatrix}
-e^{i\omega t}\lim\limits_{\varepsilon\rightarrow0}\frac{1}{(2\pi)^{3}}\int d^{3}pe^{-i\vec{p}\cdot\vec{x}}\nonumber\\
&&~~\frac{1}{(ik_{z}\mp i\varepsilon)^{2}+p^{2}}
\begin{pmatrix}
-ip_{z}-ik_{z}\pm i\varepsilon & -ip_{x}+p_{y}\\
-ip_{x}-p_{y} & ip_{z}-ik_{z}\pm i\varepsilon\\
\end{pmatrix}\nonumber\\
&&~~\times
\begin{pmatrix}
i(p_{z}-k_{z})\\
0\\
\end{pmatrix}
T,
\label{B15}
\end{eqnarray}
\end{small}
and
\begin{small}
\begin{eqnarray}
~~\chi^{(3)}&&\approx -e^{i\omega t}\lim\limits_{\varepsilon\rightarrow0}\frac{1}{(2\pi)^{3}}\int d^{3}pe^{-i\vec{p}\cdot\vec{x}}H_{02}^{-1}M_{8}\nonumber\\
&&=-e^{i\omega t}\lim\limits_{\varepsilon\rightarrow0}\frac{1}{(2\pi)^{3}}\int d^{3}pe^{-i\vec{p}\cdot\vec{x}}\nonumber\\
&&\times\frac{1}{(-2im-ik_{z}\pm i\varepsilon)^{2}+p^{2}}\nonumber\\
&&~\times\small\begin{pmatrix}
-ip_{z}-2im-ik_{z}\pm i\varepsilon & -ip_{x}+p_{y}\\
-ip_{x}-p_{y} & ip_{z}-2im-ik_{z}\pm i\varepsilon\\
\end{pmatrix}\nonumber\\
&&~\times
\begin{pmatrix}
3i(p_{z}-k_{z})-ip_{x}-ip_{y}\\
3ip_{x}-3p_{y}+(1+i)i(p_{z}-k_{z})\\
\end{pmatrix}
T.\label{B16}
\end{eqnarray}
\end{small}

\end{document}